\documentclass[sigconf]{acmart}
 
\usepackage{hyperref}
\usepackage{url}
\usepackage{multirow}
\usepackage{makecell}
\usepackage{subfigure}
\usepackage{color}
\usepackage{booktabs}
\usepackage{adjustbox}
\usepackage{color}
\usepackage{xspace}
\usepackage{todonotes}
\usepackage{paralist}
\usepackage{amsmath}
\usepackage{balance} 
\usepackage{multicol}
\usepackage{amsfonts} 
\usepackage{algorithm}
\usepackage{algorithmic}
\usepackage{paralist}

\AtBeginDocument{%
  \providecommand\BibTeX{{%
    \normalfont B\kern-0.5em{\scshape i\kern-0.25em b}\kern-0.8em\TeX}}}

\setcopyright{acmlicensed}
\copyrightyear{2018}
\acmYear{2018}
\acmDOI{XXXXXXX.XXXXXXX}

\acmConference[Conference acronym 'XX]{Make sure to enter the correct
  conference title from your rights confirmation emai}{June 03--05,
  2018}{Woodstock, NY}
\acmISBN{978-1-4503-XXXX-X/18/06}

\begin{document}

\title{Enhancing ID and Text Fusion via Alternative Training \\in Session-based Recommendation}


\author{Juanhui Li$^1$, Haoyu Han$^1$, Zhikai Chen$^1$, Harry Shomer$^1$, Wei Jin$^2$, Amin Javari$^3$, Jiliang Tang$^1$}

\affiliation{ \institution{$^1$Michigan State University, $^2$Emory University, $^3$The Home Depot \country{USA}} 
}

\email{{lijuanh1,hanhaoy1,chenzh85,shomerha,tangjili}@msu.edu, wei.jin@emory.edu, amin_javari@homedepot.com}


\newcommand{\red}[1]{\textcolor{red}{#1}}
\newcommand{\blue}[1]{\textcolor{blue}{#1}}

\newcommand{\fbt}{\texttt{FB15K-237}\xspace}
\newcommand{\nell}{\texttt{NELL-995}\xspace}

\newcommand{\jh}[1]{\textcolor{orange}{[JH: #1]}}
\newcommand{\jt}[1]{\textcolor{red}{JT:#1}}
\newcommand{\hy}[1]{\textcolor{blue}{Haoyu:#1}}
\newcommand{\wei}[1]{\textcolor{cyan}{Wei:#1}}
\newcommand{\AJ}[1]{\textcolor{violet}{Amin: #1}}

\newtheorem{finding}{Observation}

\renewcommand{\shortauthors}{Li et al.}

\begin{abstract}

  Session-based recommendation has gained increasing attention in recent years, with its aim to offer tailored suggestions based on users' historical behaviors within  sessions. 
  To advance this field, a variety of methods have been developed, with ID-based approaches typically demonstrating promising performance. 
  However, these methods often face challenges with long-tail items and overlook other rich forms of information, notably valuable textual semantic information. 
  To integrate text information, various methods have been introduced, mostly following a naive fusion framework.
  Surprisingly, we observe that fusing these two modalities does not consistently  outperform the  best single modality by following the naive fusion framework. 
  Further investigation reveals an potential imbalance issue in naive fusion, where the ID dominates and text modality is undertrained. This suggests that the unexpected observation may stem from naive fusion's failure to effectively balance the two modalities, often over-relying on the stronger ID modality.
  This insight suggests that naive fusion might not be as effective in combining ID and text as previously expected.
To address this, we propose a novel alternative training strategy  AlterRec. It separates the training of ID and text, thereby avoiding the imbalance issue seen in naive fusion. 
Additionally, AlterRec designs a novel strategy to facilitate the interaction between the two modalities, enabling them to mutually learn from each other and integrate the text more effectively.
Comprehensive experiments demonstrate the effectiveness of AlterRec in session-based recommendation. The implementation is available at \url{https://github.com/Juanhui28/AlterRec}.
\end{abstract}

\ccsdesc[500]{Do Not Use This Code~Generate the Correct Terms for Your Paper}
\ccsdesc[300]{Do Not Use This Code~Generate the Correct Terms for Your Paper}
\ccsdesc{Do Not Use This Code~Generate the Correct Terms for Your Paper}
\ccsdesc[100]{Do Not Use This Code~Generate the Correct Terms for Your Paper}

\keywords{Session Recommendation, ID-based Model, Text Information}


\received{20 February 2007}
\received[revised]{12 March 2009}
\received[accepted]{5 June 2009}

\maketitle

\section{Introduction}



In recent years, predicting the next item in user-item interaction sequences, such as clicks or purchases, has gained increasing attention~\cite{wu2019srgnn,li2017neural, pang2022hggnn, hou2022core}. This practice is prevalent across various online platforms, including e-commerce, search engines, and music/video streaming sites. These sequences are  created during user-item interactions in sessions. They encode user preferences which are dynamic and evolve over time~\cite{tahmasbi2021modeling}. For instance, a user’s interests may shift from outdoor furniture in spring to indoor tools in autumn. Moreover, in many systems, only the user's behavior history during an ongoing session is accessible. Therefore, analyzing interactions in active sessions becomes essential for real-time recommendations. This need has spurred the development of 
session-based recommendations~\cite{wu2019srgnn,hou2022core}, which utilizes the sequential patterns in a session to understand and predict the latest user preferences.



In this domain, 
 {ID-based methods}~\cite{kang2018sasrec, sun2019bert4rec,wu2019srgnn} 
 have become the predominant approach, significantly influencing the recommendation paradigm~\cite{yuan2023wheretogo, li2023upperlimit}. 
  Typically, these methods involve assigning unique ID indexes to users and items, which are then transformed into vector representations.
 Their popularity stems from their simplicity and effectiveness across various applications~\cite{li2023upperlimit}.
Despite their proven effectiveness, ID-based methods still have limitations. One drawback is their heavy reliance on the ID-based information.
They tend to overlook other forms of valuable data, notably rich text information.  This exclusion of textual data can result in less informative representations.
Such reliance can be problematic in scenarios with limited interactions between users and items. 
However, most items typically experience sparse interactions, known as long-tail items~\cite{park2008long}, which presents a challenge for these methods.


Recognizing these limitations, there has been a shift towards integrating text data for recommendations.  The surging volume of text data emphasizes the crucial role of {combing text} in various domains, such as news recommendation~\cite{li2022miner, wu2019neural} and e-commerce~\cite{jin2023amazon}.
These systems aim to accurately identify and match user preferences and interests using the available textual data. They achieve this by processing and encoding various forms of textual content, such as user reviews, product descriptions and titles, and news articles, to provide tailored recommendations. 
 Recent trends indicate an increasing reliance on language models~\cite{kenton2019bert,brown2020language,hou2022unisec,wei2023llmrec, harte2023llm2bert4rec} for extracting semantic information from textual data. It is largely due to the exceptional ability of these models to encode textual information effectively.
 This progress has sparked considerable interests in the research community, particularly in enhancing recommendation systems beyond traditional user-item interaction data~\cite{hou2022unisec,wei2023llmrec,ren2023rlmrec, yuan2023wheretogo, harte2023llm2bert4rec}. 

\begin{figure}[t]
\begin{center}
 \centering
\includegraphics[width=0.99\linewidth]{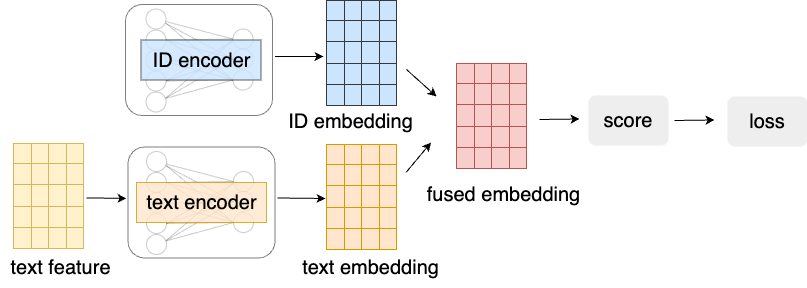} 
\caption{An illustration of a naive fusion framework.  }
  
\label{fig:naive_fusion}
\end{center}
\vspace{-0.2in}
\end{figure}

The prevailing approach in current literature for combining ID and text information typically employs a \textbf{naive fusion} framework~\cite{hou2022unisec,zhang2019fdsa, wei2023llmrec}, as shown in \figurename~\ref{fig:naive_fusion}. It involves generating embeddings from ID and text encoders, merging them to form a final embedding, and then using this for loss computation. However, our preliminary study detailed in Section~\ref{sec:pre_study}, 
reveals that the naive fusion may not be as effective as previously believed in combining ID and text information.
1) Notably, it shows that independent training on just the ID information can yield performance comparable to, or even better than, the naive fusion model. This implies that the naive fusion model may not necessarily enhance, and could potentially reduce the overall performance. This finding aligns with the studies in multi-modal learning~\cite{huang2022modality,wang2020makes,DuTLLYWYZ23}, which indicates that the fusion of multiple modalities doesn't always outperform the best single modality. 2) We further explore one naive fusion implementation as an example to have a deeper understanding of this finding.  The exploration suggests a potential imbalance issue: the model heavily relies on the ID component, while the text component appeared undertrained.  This imbalance implies that the unexpected finding might be a result of the naive fusion framework's inability to balance the contributions of the two types of information effectively, thereby hindering optimal overall performance.

 The imbalance issue identified in the naive fusion models significantly hinders the accurate integration of textual data.  Despite increased efforts to integrate textual content, these methods often fail to effectively capture essential semantic information. It results in a considerable loss of valuable information.
 This realization shifts our focus towards independent training, which does not exhibit this issue. However, independent training overlooks the potential for ID and text to provide complementary information that could be mutually learning. To address these challenges, we propose a novel \textbf{Alter}native training strategy  to combine the ID and text components for session-based \textbf{Rec}ommendation (\textbf{AlterRec}). This approach separates the training of ID and text and thereby avoiding the imbalance issue. Additionally, it goes beyond simple independent training by enabling implicit interactions between these two modalities, thereby allowing them to inform and learn from each other. More specifically, our model consists of two distinct modules: an ID uni-modal network for encoding ID information, and a text uni-modal network for processing textual data. We design an alternating update strategy, where one module learns from the other by using its generated predictions as training signal. We conduct comprehensive experiments to validate the superior effectiveness of AlterRec over a variety of baselines in real-world datasets.



\section{Related Work}

\subsection{ID-based Methods}
These methods convert each user or item into a vector representation using unique ID indices. Traditionally,
deep neural networks serve as encoders in this context. For example, GRU4REC~\cite{HidasiKBT15} employs a recurrent neural network (RNN) to analyze user-item interaction sequences. 
More recent advancements have seen the adoption of sophisticated architectures as encoders. 
For instance, SASRec~\cite{kang2018sasrec} implements a Transformer encoder with self-attention to delineate user preferences within sequences. BERT4Rec~\cite{sun2019bert4rec} uses the BERT model, incorporating a cloze objective to model user behaviors. Another approach employs GNNs as encoders~\cite{wu2019srgnn, pang2022hggnn}. For example, SR-GNN~\cite{wu2019srgnn} constructs session-specific graphs and introduces a gated GNN to capture complex item transitions. Similarly, HG-GNN~\cite{pang2022hggnn} builds a heterogeneous user-item graph to elucidate user-item transition patterns across multiple sessions.  
However, these methods overlook additional valuable text information, potentially leading to less informative representations.


\subsection{Text-Integrated Methods}.
 These methods combine the text information to perform recommendations. 
 For instance, 
 FDSA~\cite{zhang2019fdsa} leverages Word2Vec~\cite{mikolov2013efficient} for semantic representation 
 and integrate both ID-based and text-based embeddings via the concatenation operation.
S³-Rec~\cite{zhou2020s3rec} utilizes an embedding matrix for learning text embeddings, combining textual information through various self-supervised tasks. More recent developments involve the use of language models for text-based embedding due to their advanced text modeling capabilities. UniSRec~\cite{hou2022unisec} employs the BERT model for text embedding 
and combines the ID-based by summing them.
LLM2BERT4Rec~\cite{harte2023llm2bert4rec} designs to use the text feature extracted by the large language model as the initialization of the item ID embeddings.
RLMRec~\cite{ren2023rlmrec} and LLMRec~\cite{wei2023llmrec} both use large language models (LLMs) for generating user/item profiles and encoding them into semantic representations. 
RLMRec aligns ID-based embedding with textual information using contrastive and generative loss, whereas LLMRec
integrates the text-based embeddings via summation.
Among the methods discussed, the majority follows the naive fusion framework~\cite{zhang2019fdsa,hou2022unisec,wei2023llmrec}, which may not effectively incorporate text information as identified in Section~\ref{sec:pre_study}.

\begin{figure*}[t]
\begin{center}
 \centerline{
{\subfigure[NFRec]
{\includegraphics[width=0.3\linewidth]{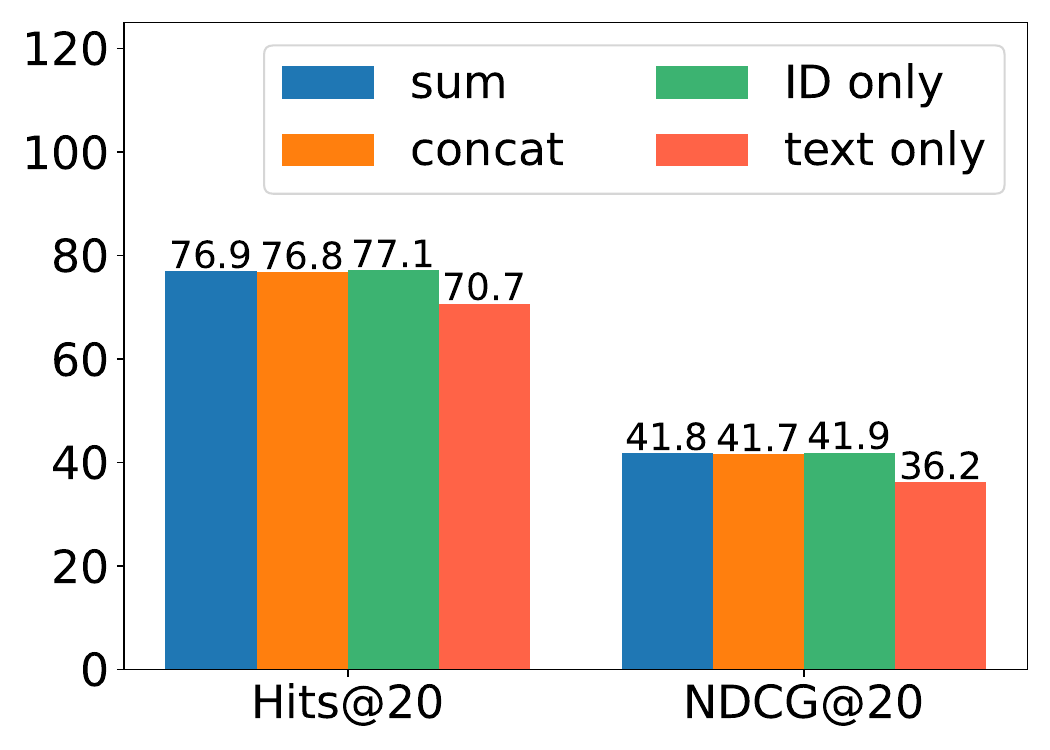} }}

{\subfigure[UnisRec]
{\includegraphics[width=0.3\linewidth]{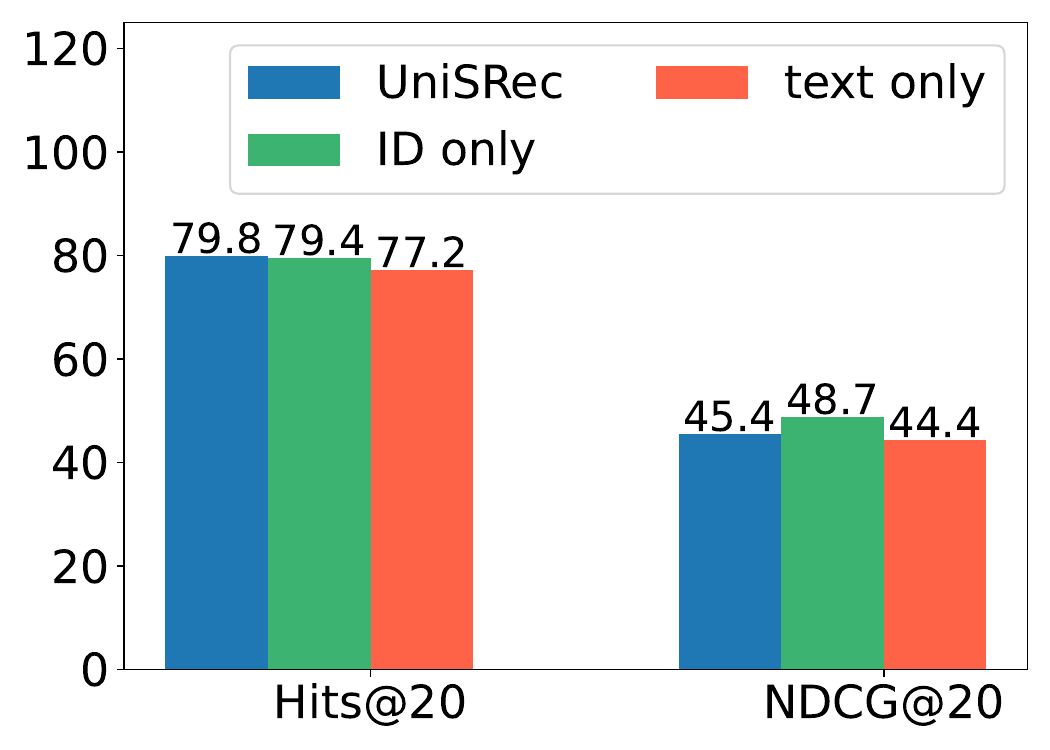} }}

{\subfigure[FDSA]
{\includegraphics[width=0.3\linewidth]{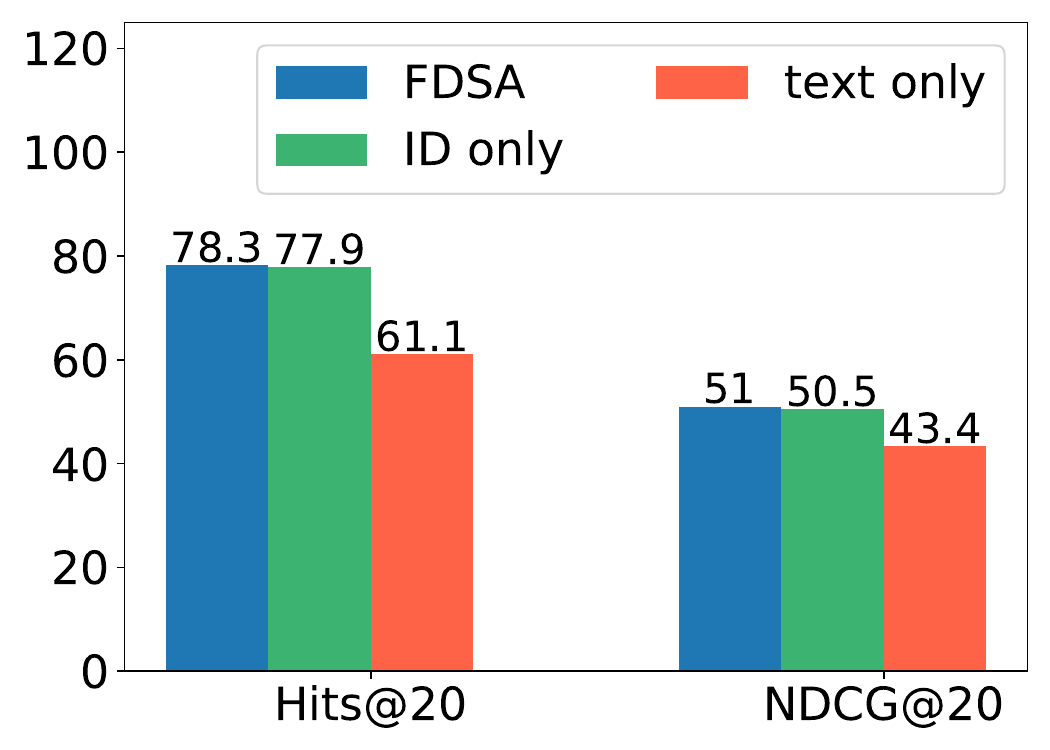} }}

}
\vspace{-0.2in}
\caption{Session-based recommendation results (\%) on the Amazon-French dataset. We compare the  models combing ID and text against models trained independently on either ID or text information alone }

\label{fig:prili_spanish}
\end{center}
\vspace{-0.2in}
\end{figure*}

\subsection{Multi-modal Learning}
The multi-modal learning paradigm, which integrates various data types like text, video, and audio~\cite{liang2021multibench}, aims to enhance overall performance by leveraging the strengths of each modality. However, recent studies have found that fusing multi-modal data does not always outperform best single modality~\cite{peng2022balanced,wang2020makes,huang2022modality,DuTLLYWYZ23,wu2022characterizing,du2023uni}. They study the issue from different angles.
For instance, G-Blend~\cite{wang2020makes} identifies an overfitting issue and varying convergence rates across modalities, proposing a gradient-blending method for optimal modality integration. OGM-GE~\cite{peng2022balanced}  suggests that a dominant strong modality might lead to imbalanced optimization and introduces an on-the-fly gradient modulation strategy to adjust each modality's gradient.
\citet{huang2022modality} theoretically demonstrate a competition between modalities, where the losing modality fails to be adequately utilized. 

\section{Preliminaries}

\subsection{Session-based Recommendation}

Consider a set of users and items, denoted as $\mathcal{U}$ and $\mathcal{V}$, respectively.  
Let $\mathcal{S}$ denote the set of user-item interaction sequences (or sessions). We use $ \mathbf{s} = \{ s_1, s_2,, ..., s_n\} \in \mathcal{S}$ to represent one of the sequences, where $s_i \in \mathcal{V}$ is the $i$-th item interacted with by the same user in session $\mathbf{s}$, and $n$ is the total number of interactions. The number of users and items are represented by  $|\mathcal{U}|$ and $|\mathcal{V}|$, respectively. Each item $i$ in $\mathcal{V}$ is associated with text information, such as product descriptions, titles, or taxonomies, denoted by $t_i = \{ w_1, w_2, ..., w_c\}$. Here, each word  $w_j$ belongs to a shared vocabulary, and $c$ represents the truncated length of the text. Given an item sequence,
the objective of session-based recommendation is to predict the next item in the current sequence. Formally, {this involves generating a ranking list $\mathbf{y}_s$ for all candidate items , where $\mathbf{y}_\mathbf{s} = [y_{\mathbf{s},1}, ..., y_{\mathbf{s},|\mathcal{V}|}]$ and each $y_{\mathbf{s},i}$ is a score indicating the likelihood of item $i$ being the next interacted item given a session $\mathbf{s}$.} 

\noindent \textbf{The Naive Fusion Framework}.
In the realm of session-based recommendation tasks, ID-based and text-based information can be combined to potentially improve the overall performance.
 The majority of existing methods employ a naive fusion approach combined with a joint training strategy~\cite{hou2022unisec,zhang2019fdsa,wei2023llmrec}.
 We present the illustration of this framework in \figurename~\ref{fig:naive_fusion}. Specifically, this approach involves generating two types of embeddings $\mathbf{X}^{ID}$ and $\mathbf{X}^{text}$ 
using ID and text encoders, respectively. These two  embeddings can be item-level or session-level.
The embeddings are then merged to form a  final embedding, denoted as $\mathbf{Z}$, through methods such as summation or concatenation.  This final embedding is used to calculate a relevance score between a given session and the candidate item.  
This score estimates the likelihood of the item being the next choice in the session. Then the score  is used to compute of the loss which optimizes the entire framework. 
Throughout this paper, the term \textbf{naive fusion} is used to refer to this framework. Notably, existing methods such as UniSRec~\cite{hou2022unisec}, FDSA~\cite{zhang2019fdsa}, and LLMRec~\cite{wei2023llmrec}  follow this approach.


\vspace{-0.1in}
\subsection{Preliminary Study}
\label{sec:pre_study}


In this subsection, motivated by multi-modal learning~\cite{wang2020makes,huang2022modality, peng2022balanced}, we conduct a preliminary study to investigate potential challenges in combining ID and text information for session-based recommendation. This study could motivate more effective integration strategy for these two types of information.

 In the naive fusion framework, the ID and text can be treated as two different types of modality that work together to improve the overall performance.  However,  studies in the multi-modal learning ~\cite{wang2020makes,DuTLLYWYZ23,huang2022modality, peng2022balanced} reveals \textbf{a phenomenon: fusing two modalities does not usually outperform the best single modality trained independently.} 
In other words, combining modalities may not enhance, and could potentially reduce, overall performance. {Various studies have focused on this phenomenon, offering analysis from different perspectives, such as greedy learning~\cite{wu2022characterizing}, modality competition~\cite{huang2022modality} and modality laziness~\cite{du2023uni}.
}
To effectively merge ID and text information for session-based recommendations, we conduct an investigation to first verify the presence of this phenomenon and then explore its underlying causes. Further details will be provided in the following subsections.

\subsubsection{Naive Fusion vs. Independent Training}
\label{sec:joint_vs_independent}
{To examine if the phenomenon mentioned above exists in session-based recommendation,  
this investigation aims to compare the performance of 
naive fusion models including our implementation named NFRec,  UniSRec~\cite{hou2022unisec} and FDSA~\cite{zhang2019fdsa} against their corresponding two single modality models (ID and text) that are trained independently. }  
To ensure a fair comparison, we employ the same ID/text encoder, scoring function, and loss function  across both naive fusions and independent training frameworks. These methods have different implementation on the key components. 
More details are given in  Appendix~\ref{sec:implementation_naive_fusion}.
{The results on  Amazon-French (detailed in Section~\ref{sec:datasets}) are presented in Figure~\ref{fig:prili_spanish} }, where ``ID only'' and ``text only'' denote the respective ID and text only models that are trained independently.  Furthermore, in  Figure~\ref{fig:prili_spanish}(a), ``sum'' and  ``concat'' represent our naive fusion implementations using summation and concatenation respectively to combine the two types of information.  Due to space limit, we present additional results on another dataset Homedepot in Appendix~\ref{sec:app_additional_result}, which shows similar phenomenon.
We employ two widely used metrics Hits@20 and NDCG@20, where higher scores indicate better performance.
We have the following observations:
\begin{finding}\label{find:ID} 
 Training solely with ID information independently can often achieve performance comparable to, or even better than, naively fusing  both ID and text. This
 indicates   the ineffectiveness of naive fusion as a method for combining ID and text.
\end{finding}
\begin{finding}\label{find:text} 
The  text only model generally results in the worst performance,  often exhibiting a substantial performance gap compared to the ID only approach.

\end{finding}

The first observation aligns with findings in multi-modal learning studies~\cite{peng2022balanced,huang2022modality,wang2020makes}, indicating a similar phenomenon. This suggests that the integration of ID and text information is not as effective as expected in the session-based recommendation. To gain a more comprehensive understanding of this issue, we will delve into NFRec, which are elaborated upon in the following subsection.

\begin{figure}[t]
\begin{center}
 \centerline{


{\subfigure[Amazon-French]
{\includegraphics[width=0.5\linewidth]{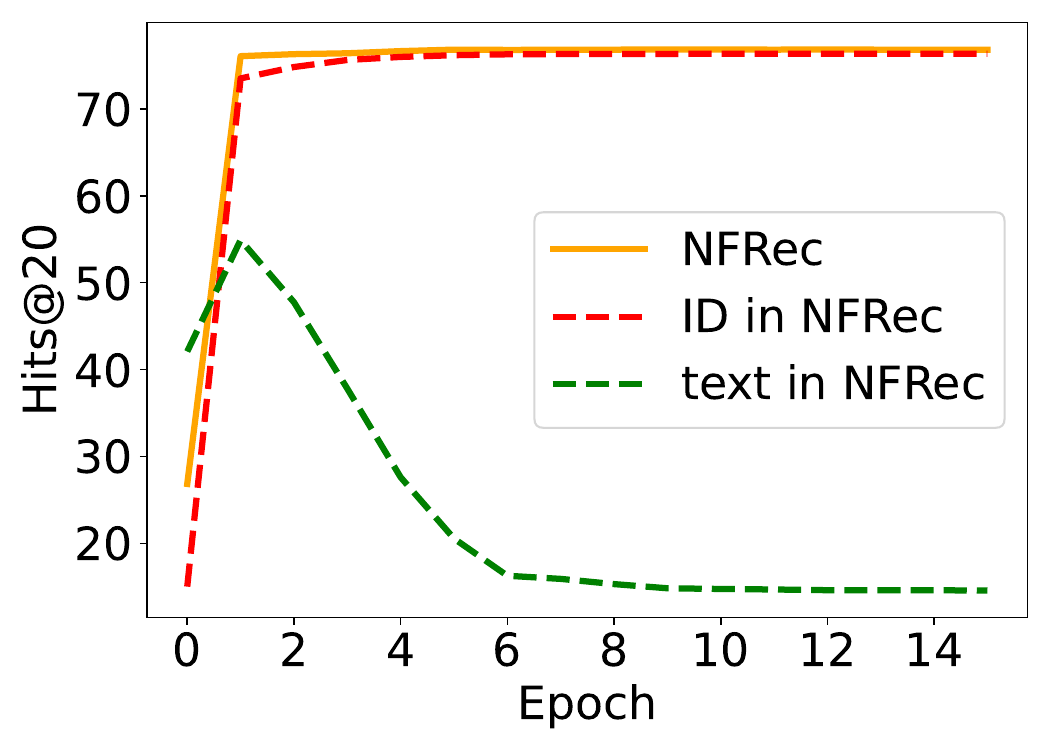} }}

{\subfigure[Amazon-French]
{\includegraphics[width=0.5\linewidth]{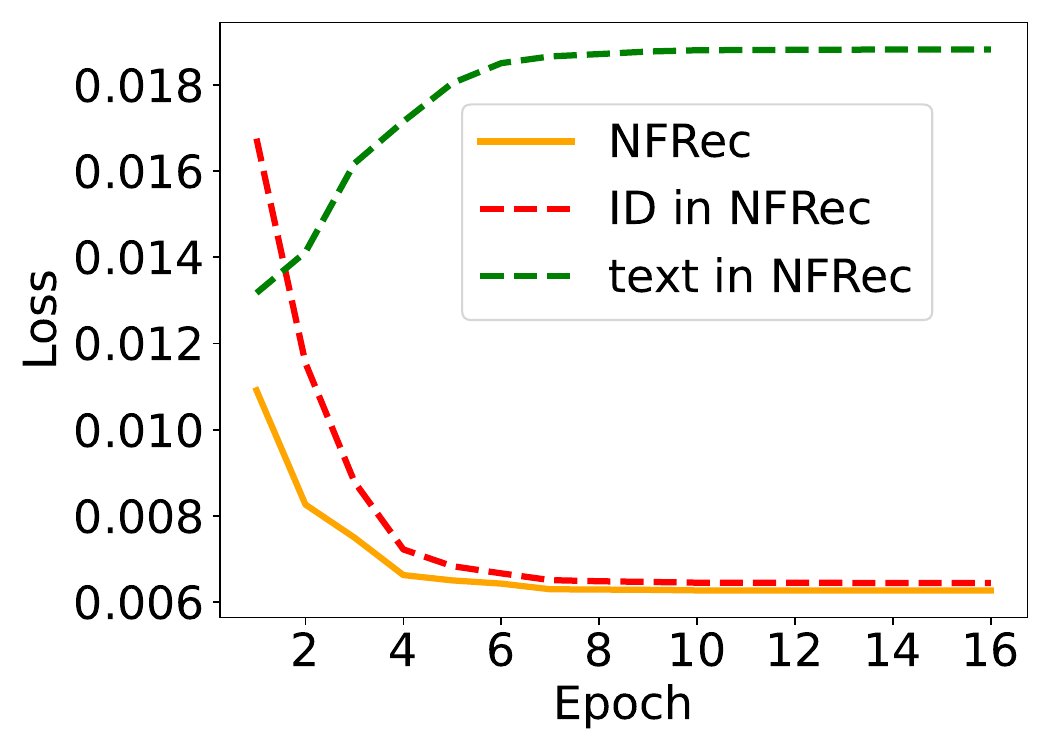} }}

}
\vspace{-0.1in}
\caption{Test performance in terms of Hits@20 (\%) and  training loss comparison on  the Amazon-French dataset.  }

\label{fig:understand}
\vspace{-0.35in}

\end{center}

\end{figure}

\subsubsection{Exploration of  NFRec}
\label{sec:explore_naive_fusion}

In our exploration,  we aim to understand how ID and text components perform under NFRec.
{It will provide insights into why combining these components in a naive fusion framework does not yield the expected improvement.
To this end, we take NFRec applying {concatenation} for fusing ID and text information as one example. The NFRec can be conceptually divided into two segments: the ID component and the text component.}  
Additional details on this division are available in the Appendix~\ref{sec:app_explore_naive_fusion}.
%
Test performance and training loss across each epoch on the Amazon-French are presented in Figure~\ref{fig:understand}. For clarity, we label the two components as "ID in NFRec" and "text in NFRec" represented by red and green dashed lines, respectively.
It is important to note that at epoch 0, all models are in a randomly initialized and  untrained state. Additional  results on the Homedepot dataset are given in Appendix~\ref{sec:app_additional_result}.



Figure~\ref{fig:understand} reveals a significant imbalance issue in NFRec: 
 the performance and loss  of the ID component are almost overlapping with those of NFRec.
This indicates a heavy reliance on the ID component in NFRec,  where the ID dominates the overall performance and loss, and the text component has limited contributions.
This suggests that the first observation in Section~\ref{sec:joint_vs_independent} may stem from the nature of naive fusion. 
{
Specifically, it appears incapable of balancing the modalities to achieve optimal overall performance and tends to overly depend on the stronger ID modality (as noted in the second observation in Section~\ref{sec:joint_vs_independent}).
}
Supporting this hypothesis is from various studies~\cite{peng2022balanced,huang2022modality,wang2020makes, wu2022characterizing} in multi-modal learning which offer empirical and theoretical insights from different angles, such as greedy learning~\cite{wu2022characterizing}, modality competition~\cite{huang2022modality} and modality laziness~\cite{du2023uni}.
Further investigation to identify more concrete causes of this phenomenon is designated as one future work.

{
The analysis above reveals an imbalance issue deriving from naive fusion, suggesting it may not be an effective method for combining ID and text information. 
To address this, we propose a novel approach involving the alternate training of ID and text information. This method is designed to encourage implicit interactions between the ID and text. The details of this proposed framework are elaborated in the following section.
}


\begin{figure}[t]
\begin{center}
 \centering
\includegraphics[width=1\linewidth]{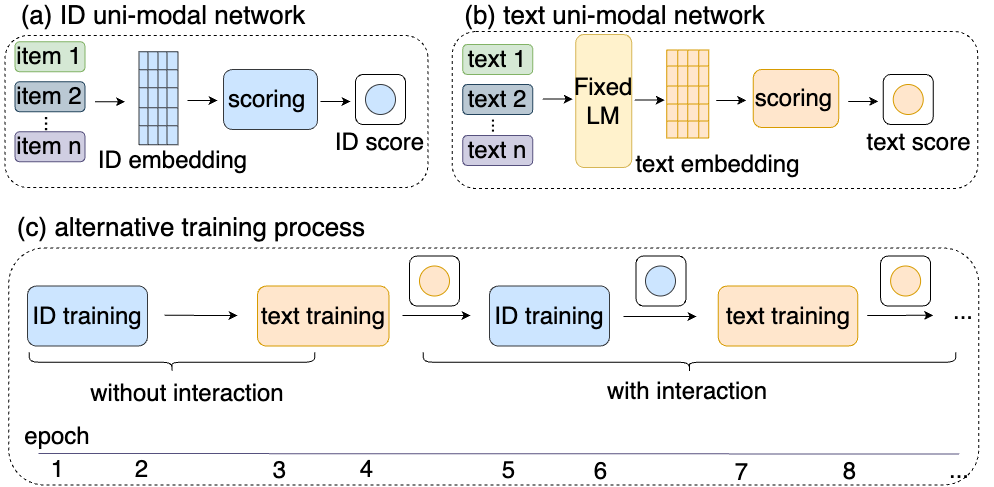} 
\vspace{-0.25in}
\caption{    
An Overview of AlterRec. (a), (b): two key components – the ID and text uni-modal networks. (c): These networks are trained alternately, learning from each other through predictions generated by the other network.
}
  
\label{fig:AlterRec}
\end{center}
\vspace{-0.2in}
\end{figure}

\section{Framework}

Having identified the potential imbalance issue with the naive fusion framework in the previous section, we explore to combine ID and text information by training them separately. However, simply training ID and text independently may not fully exploit their potential to provide complementary information.
To address these challenges, we introduce AlterRec, a novel alternative training method. Figure~\ref{fig:AlterRec} provides an overview of AlterRec. The model comprises two key components: the ID and text uni-modal networks, designed to capture ID and semantic information respectively. We employ the predictions from one network as training signals for the other, facilitating interaction and mutual learning through these predictions.
AlterRec  separates the training of ID and text, effectively avoiding the imbalance issue, as shown in section~\ref{sec:ablation_study}. Moreover, it goes beyond independent training by facilitating interaction between ID and text components, enabling them to learn mutually beneficial information from each other and incorporate the text more effectively. Next, we provide further details of these components.




\subsection{ID and Text Uni-modal Networks }

The ID and text unimodal networks share similar architectures.  Each has respective ID/text encoders to generate ID and text embeddings. Based on these embeddings, a scoring function is adopted to calculate  the relevance between  given session and candidate items. We first introduce two encoders, and then use the ID embedding as an example to illustrate how to define  the scoring function. 

\subsubsection{ID Encoder}
\label{sec:id_encoder}
The ID encoder in our model is designed to create a unique embedding for each item based on its ID index. This is achieved using an ID embedding matrix $\mathbf{X} \in \mathbb{R}^{|\mathcal{V}|\times d} $, where $d$ represents the size of the embedding. Each row, $\mathbf{X}_i$, corresponds to the ID embedding of item $i$. Notably, this matrix is a learnable parameter within the network and updated during the optimization. 
   

\subsubsection{Text Encoder} 
\label{sec:text_encoder}
The text encoder in our model is designed to extract textual information from items. Leveraging the advanced language modeling capabilities, we utilize the Sentence-BERT~\cite{reimers2019sentence} in this work. For each item 
$i$, represented by a text sequence $t_i = \{ w_1, w_2, ..., w_c\}$, the  Sentence-BERT processes the input sentence to generate token embeddings. These embeddings are then aggregated using a pooling layer to form a comprehensive embedding for the entire sentence. And we further use a MLP~\cite{delashmit2005recent} to transform the embeddings generated from the  Sentence-BERT into a  $d-$dimensional matrix.
Formally: 
\begin{equation}
\label{eq:text_embedding}
    \mathbf{H}_i = \text{MLP}(\text{SBERT}( w_1, w_2, ..., w_c) ), 
\end{equation}
where  $\mathbf{H} \in \mathbb{R}^{|\mathcal{V}|\times d}$ and
each row $\mathbf{H}_i$ corresponds to the text embedding of item $i$. 
Considering practical constraints, we fix the language model which isn't updated during the optimization process due to the high training cost.



\subsubsection{Scoring Function}
\label{sec:id_score}
In the context of session-based recommendation, our objective is to predict the next item in a sequence of items interacted with by the same user, denoted as  $\mathbf{s} = \{ s_1, s_2,..., s_n \}$. To accomplish this, we generate a prediction score, $y_{\mathbf{s},j}$, for each candidate item  $j$. These scores are then used to rank all candidate items, with the top-ranked item predicted as the next item. The process begins with obtaining the session embedding $\mathbf{q}_{\mathbf{s}} \in \mathbb{R}^{d}$ for session $\mathbf{s}$. This embedding encodes the user's interaction behavior within that session. The relevance between the session embedding and each candidate item's embedding is calculated and used as the score for that item.  Next, we introduce two approaches to generate the session embedding, and use the ID embedding as an example since it's similar for the text embedding.

\noindent \textbf{Session Embedding via Mean Function}. To encode the interaction information within session $\mathbf{s}$, we can use a mapping function, $\mathbf{q}_{\mathbf{s}} = g(\mathbf{s}, \mathbf{X}) $ to operate on the item ID embeddings. A simple yet effective approach is to calculate the mean of the item embeddings within the session. Formally, this is represented as:
\begin{equation}
\label{eq:id_mean}
    g_{mean}(\mathbf{s}, \mathbf{X}) = \frac{1}{|\mathbf{s}|}\sum_{s_i \in \mathbf{s}} \mathbf{X}_{s_i}
\end{equation}
Here, $|\mathbf{s}|$ denotes the length of the session $\mathbf{s}$.

\noindent \textbf{Session Embedding via Transformer}. To capture the item-item transition patterns within the same session, we employ the transformer architecture~\cite{vaswani2017attention}, which utilizes self-attention to weight the relative influence of different items.  Formally: 
\begin{equation}
    \hat{\mathbf{E}} = \text{Transformer}(\mathbf{X}_{s_1}, \mathbf{X}_{s_2}, \mathbf{X}_{s_n})
\end{equation}
where $\hat{\mathbf{E}}\in \mathbb{R}^{n\times d}$ is the refined embedding after the Transformer. Given the sequential nature of session, we employ a masking matrix in self attention to prevent the model from accessing future information~\cite{zhang2019fdsa}. Due to the space limit, more details of the masking matrix is presented in Appendix~\ref{sec:app_masking_matrix}.
Consequently, the embedding of the last item in the sequence is used as the session embedding, as it encapsulates information from all items in the current session. 
\begin{equation}
\label{eq:id_transformer}
g_{Trans}(\mathbf{s}, \mathbf{X}) = \hat{\mathbf{E}}_{n}
\end{equation}
where $\hat{\mathbf{E}}_{n}$ is the $n$-th row of $\hat{\mathbf{E}}$.

\noindent \textbf{Session\&Item Relevance}.
The session embedding can be derived using either $\mathbf{q}_{\mathbf{s}} =g_{mean}(\mathbf{s}, \mathbf{X})$ or $\mathbf{q}_{\mathbf{s}} = g_{Trans}(\mathbf{s}, \mathbf{X})$, based on their empirical performance. We determine the relevance between the session and each candidate item using vector multiplication:
\begin{equation}
\label{eq:id_score}
y^{ID}_{\mathbf{s},i} = \mathbf{X}^T_i\mathbf{q}_{\mathbf{s}} 
\end{equation} 
Similarly, we can get the score $y^{Text}_{\mathbf{s},i}$ based on the text embedding:
\begin{equation}
\label{eq:text_score}
    y^{Text}_{\mathbf{s},i} = \mathbf{H}^T_i \hat{\mathbf{q}}_{\mathbf{s}} 
\end{equation}
where $\hat{\mathbf{q}}_{\mathbf{s}} = g_{mean}(\mathbf{s}, \mathbf{H})$ or $\hat{\mathbf{q}}_{\mathbf{s}} = g_{Trans}(\mathbf{s}, \mathbf{H})$.


\subsection{Alternative Training }
\label{sec:negative_sample}
The ID and text data offer different types of information. Our goal is to facilitate their interaction, enabling mutual learning and thereby enhancing overall performance. To this end, we propose an alternative training strategy to use predictions from one uni-modal network to train the other network. These predictions encode information of one modality, allowing one network to learn information from the other. We leverage the predictions from one modality to the other in two aspects. First, we select top-ranked items as augmented positive training samples. These items with top scores are likely very relevant to the current session from the perspective of one modality that could provide more training signals for the other modality especially for items with fewer interactions. Second, we choose other high-scored items as negative samples. These items are ranked higher but not the most relevant ones for one modality and we aim to force the other modality to distinguish them from positive samples. Such negative samples are much harder to be distinguished compared to those from traditional random sampling~\cite{rendle2012bpr}. Thus, we refer to them as hard negative samples in this work. 
For illustrative purposes, we will use the predictions from the ID uni-modal network to train the text uni-modal network as an example. 


\noindent \textbf{Hard Negative Samples.} We first generate predictions from the ID uni-modal network for a given session $\mathbf{s}$. Then  we rank the scores of all candidate items in descending order:
\begin{equation}
    r^{ID}_{\mathbf{s}} = \text{argsort}(y^{ID}_{\mathbf{s},1}, y^{ID}_{\mathbf{s},2}, ..., y^{ID}_{\mathbf{s}, |\mathcal{V}|})
\end{equation}
Here, $r^{ID}_{\mathbf{s}}$ 
denotes the sequence of ID indices corresponding to the sorted scores. 

We select items ranked from $k_{1}$ to $k_{2}$, represented as  $r^{ID}_{\mathbf{s}} [k_{1}:k_{2}]$, as the {hard negative samples} for training the text uni-modal network. It enables the text uni-modal network to learn from the patterns identified by the ID uni-modal network.
These hard negatives play a crucial role in defining the loss function. For a given session $\mathbf{s}=\{s_1, s_2, ..., s_n \}$ with $s_t$ as the target item, we use the  cross entropy  as the loss function by following most of the related works~\cite{hou2022unisec,wu2019srgnn,pang2022hggnn}:
\begin{equation}
\label{eq:text_loss}
    L^{text} = -\sum_{\mathbf{s}\in \mathcal{S}}   \log (f(y^{text}_{\mathbf{s}, s_{t}}))  
\end{equation}
where $f$ is the Softmax function applied over the target item $s_t$ and the negative samples in $r^{ID}_{\mathbf{s}} [k_{1}:k_{2}]$. 

Similarly, we can use the hard negative sample  $r^{text}_{\mathbf{s}} [k_1:k_2] $ derived from the text uni-modal network to train the ID uni-modal network. It's  obtained by sorting the scores and identifying the ranking ID index $  r^{text}_{\mathbf{s}} = \text{argsort}(y^{text}_{\mathbf{s},1}, y^{text}_{\mathbf{s},2}, ..., y^{text}_{\mathbf{s},|\mathcal{V}|})$.  We 
define the loss function for the ID uni-modal network as follows:
\begin{equation}
\label{eq:id_loss}
    L^{ID} = -\sum_{\mathbf{s}\in \mathcal{S}}   \log (f(y^{ID}_{\mathbf{s}, s_{t}})) 
\end{equation}
where the the Softmax function is applied over the target item $s_t$ and the negative samples in $r^{text}_{\mathbf{s}} [k_{1}:k_{2}]$.


\noindent \textbf{Positive Sample Augmentation}. 
To train the text uni-modal network, we utilize  $r^{ID}_\mathbf{s} [1:p] $  as additional positive samples which serves as ground-truth target items.  
Similarly, $r^{text}_{\mathbf{s}} [1:p] $ is used as supplementary positive samples for training the ID uni-modal network. Typically, we set $p < k_1$. Accordingly, the loss function in Eq.~(\ref{eq:text_loss}) and (\ref{eq:id_loss}) is modified as follows:
\begin{equation}
    \label{eq:text_loss_aug}
    L_{a}^{text} = -\sum_{\mathbf{s}\in \mathcal{S}} \bigg(  \log (f(y^{text}_{\mathbf{s}, s_{t}})) + \beta*\sum_{s_k \in r^{ID}_\mathbf{s} [1:p]}\log (f(y^{text}_{\mathbf{s}, s_{k}}))\bigg)
\end{equation}
\begin{equation}
\label{eq:id_loss_aug}
    L_{a}^{ID} = -\sum_{\mathbf{s}\in \mathcal{S}}  \bigg( \log (f(y^{ID}_{\mathbf{s}, s_{t}})) +\beta*\sum_{s_k \in r^{text}_\mathbf{s} [1:p]}\log (f(y^{ID}_{\mathbf{s}, s_{k}})) \bigg)
\end{equation}
Here, $\beta$ is a parameter to adjust the importance of the augmented samples. Note that within each network, these augmented samples are paired with the same corresponding hard negative samples as the target item $s_t$. 

\noindent \textbf{ Training Algorithm}.
This algorithm focuses on facilitating the interaction between two networks, and we  use the \figurename~\ref{fig:AlterRec}(c) as a more straightforward illustration. 
The training process consists of two stages. \textbf{1) Initially}, 
 due to the lower quality of the learned embeddings, we don't employ interaction between  two networks. Thus, we apply random negative samples during the first $m_{random}$ epochs. This involves replacing the hard negatives in Eq.~(\ref{eq:text_loss}) and Eq.~(\ref{eq:id_loss})   with randomly selected negatives with equal number.
\textbf{2) Subsequently}, we shift to training with hard negatives. We start by training the ID uni-modal network using hard negatives derived from the text uni-modal network.
After  $m_{gap}$ 
  epochs, the training focus shifts to the text uni-modal network, which is  trained using hard negatives from the ID uni-modal network.
  Following another  $m_{gap}$ 
  epochs, we resume training the ID uni-modal network and repeat this alternating process. This approach ensures that each network  continually learns from the other, thereby potentially improving overall performance. More details of the training process are given in Appendix~\ref{sec:app_algorithm}.

Upon convergence of both networks, we generate a final relevance score by combining the relevance scores from each network and weighting their contributions. This score is used during the \textbf{inference stage} and is defined as:
\begin{equation}
\label{eq:final_score}
    y_{\mathbf{s},i} = \alpha* y^{ID}_{\mathbf{s},i} + (1-\alpha)*y^{text}_{\mathbf{s},i}
\end{equation}
Here, $ y_{\mathbf{s},i}$ is the final score for the candidate item $i$ given  session $\mathbf{s}$, and $\alpha$ is a pre-defined parameter.





\begin{table*}[!ht]
\centering
\caption{Performance Comparison (\%) on the Homdepot and three Amazon-M2 datasets.  All reported results are mean and standard deviation
over three seeds. 
The best results are highlighted in bold, and the second-best results are underlined.   }
\label{table:mainresult1}
\begin{adjustbox}{width =1 \textwidth}
\begin{tabular}{c|cccc|cccc}
\toprule
 &  \multicolumn{4}{c|} {Homedepot}&  \multicolumn{4}{c} {Amazon-Spanish}   \\

 & Hits@10 & Hits@20 & NDCG@10 & NDCG@20 & Hits@10 & Hits@20 & NDCG@10 & NDCG@20 \\
  \midrule
SASRec & 33.58 ± 0.27 & 40.93 ± 0.14 & 18.23 ± 0.06 & 20.09 ± 0.06 & 70.95 ± 0.32 & 80.46 ± 0.32 & 44.88 ± 0.33 & 47.29 ± 0.34 \\
BERT4Rec & 26.06 ± 0.26 & 31.85 ± 0.45 & 15.61 ± 0.3 & 17.08 ± 0.35 & 64.6 ± 0.13 & 74.0 ± 0.33 & 44.6 ± 0.16 & 46.98 ± 0.18 \\
SRGNN & 30.09 ± 0.07 & 36.0 ± 0.19 & 15.31 ± 0.13 & 15.73 ± 0.13 & 67.02 ± 0.29 & 76.37 ± 0.12 & 46.75 ± 0.33 & 49.12 ± 0.26 \\
HG-GNN & 33.17 ± 0.13 & 40.72 ± 0.20 & 18.27 ± 0.49 & 20.19 ± 0.51 & N/A & N/A & N/A & N/A \\
CORE& 37.04 ± 0.11	&44.73 ± 0.06 & 19.86 ± 0.14&	21.81 ± 0.14& 71.83 ± 0.15	&81.14 ± 0.17 & 41.05 ± 0.06	&43.41 ± 0.08 \\
UnisRec (FHCM) & {36.03 ± 0.12} & {43.67 ± 0.06} & 20.14 ± 0.79 & 22.08 ± 0.77 & 72.15 ± 0.01 & 81.3 ± 0.02 & 44.87 ± 0.1 & 47.2 ± 0.1 \\
UnisRec & 34.56 ± 0.23 & 42.19 ± 0.16 & 19.01 ± 0.08 & 20.92 ± 0.08 & {72.33 ± 0.06} & 81.42 ± 0.16 & 45.51 ± 0.05 & 47.82 ± 0.06 \\
FDSA & 32.1 ± 0.34 & 39.11 ± 0.2 & {20.44 ± 0.1} & {22.21 ± 0.06} & 70.55 ± 0.24 & 79.84 ± 0.08 & 49.83 ± 0.15 & 52.18 ± 0.13 \\
S³-Rec & 26.69 ± 0.1 & 33.04 ± 0.34 & 16.01 ± 0.14 & 17.62 ± 0.11 & 69.61 ± 0.4 & 78.85 ± 0.62 & 47.25 ± 0.45 & 49.6 ± 0.4 \\
LLM2SASRec & 34.12 ± 0.29 & 42.13 ± 0.18 & 18.69 ± 0.26 & 20.72 ± 0.22 & 71.55 ± 0.06 & 80.68 ± 0.12 & 48.45 ± 0.15 & 50.77 ± 0.17 \\
LLM2BERT4Rec & 29.51 ± 0.35 & 37.3 ± 0.33 & 16.25 ± 0.3 & 18.22 ± 0.3 & 66.47 ± 0.2 & 76.95 ± 0.27 & 40.29 ± 0.32 & 42.95 ± 0.35 \\
\midrule
AlterRec& \underline{38.25 ± 0.14}	& \underline{46.31 ± 0.11} & \underline{20.72 ± 0.06}	&\textbf{22.76 ± 0.06} & \underline{72.41 ± 0.17} & \textbf{81.49 ± 0.09} & \textbf{50.59 ± 0.14} & \textbf{52.9 ± 0.12}\\
AlterRec\_{aug} & \textbf{38.46 ± 0.1} & \textbf{46.37 ± 0.08} & \textbf{20.74 ± 0.05} & \underline{22.75 ± 0.03} & \textbf{72.47 ± 0.19}&	\underline{81.45 ± 0.04}  & \underline{50.58 ± 0.02}	& \underline{52.86 ± 0.05}\\
\midrule
&  \multicolumn{4}{c|} {Amazon-French}&  \multicolumn{4}{c} {Amazon-Italian}   \\

 & Hits@10 & Hits@20 & NDCG@10 & NDCG@20 & Hits@10 & Hits@20 & NDCG@10 & NDCG@20 \\
 \midrule
SASRec & 69.2 ± 0.15 & 78.4 ± 0.1 & 44.89 ± 0.43 & 47.23 ± 0.44 & 68.25 ± 0.08 & 78.37 ± 0.06 & 43.24 ± 0.18 & 45.81 ± 0.18 \\
BERT4Rec & 63.01 ± 0.11 & 72.47 ± 0.15 & 43.84 ± 0.04 & 46.24 ± 0.07 & 62.24 ± 0.26 & 72.38 ± 0.13 & 42.42 ± 0.15 & 44.99 ± 0.11 \\
SRGNN & 65.61 ± 0.09 & 74.93 ± 0.09 & 46.27 ± 0.1 & 48.64 ± 0.08 & 65.62 ± 0.26 & 75.2 ± 0.15 & 44.85 ± 0.17 & 47.28 ± 0.15 \\
CORE & 69.93 ± 0.02	& 79.32 ± 0.1 & 39.4 ± 0.05 &	41.79 ± 0.07 & 69.42 ± 0.12	& 79.4 ± 0.1 & 39.27 ± 0.05	&41.8 ± 0.05\\
UniSRec (FHCM) & 70.35 ± 0.04 & 79.73 ± 0.13 & 43.99 ± 0.12 & 46.37 ± 0.1 & 69.95 ± 0.06 & \underline{79.84 ± 0.07} & 42.97 ± 0.18 & 45.48 ± 0.2 \\
UniSRec & 70.54 ± 0.09 & 79.74 ± 0.03 & 44.5 ± 0.06 & 46.84 ± 0.06 & \underline{69.99 ± 0.07} & 79.63 ± 0.03 & 43.42 ± 0.08 & 45.87 ± 0.06 \\
FDSA & 68.94 ± 0.29 & 78.16 ± 0.13 & 48.62 ± 0.11 & 50.96 ± 0.08 & 67.88 ± 0.07 & 77.97 ± 0.11 & 47.04 ± 0.11 & 49.6 ± 0.12 \\
S³-Rec & 62.82 ± 1.78 & 72.85 ± 1.01 & 40.84 ± 2.57 & 43.39 ± 2.37 & 60.6 ± 2.92 & 71.67 ± 2.19 & 37.88 ± 3.42 & 40.69 ± 3.22 \\
LLM2SASRec & 70.01 ± 0.1 & 79.15 ± 0.08 & 48.13 ± 0.08 & 50.45 ± 0.11 & 69.2 ± 0.14 & 79.11 ± 0.06 & 46.22 ± 0.39 & 48.73 ± 0.4 \\
LLM2BERT4Rec & 65.48 ± 0.02 & 75.91 ± 0.08 & 39.8 ± 0.16 & 42.45 ± 0.15 & 64.88 ± 0.44 & 75.9 ± 0.14 & 31.23 ± 0.26 & 32.0 ± 0.24 \\
\midrule
AlterRec  & \underline{70.61 ± 0.03}	&\underline{79.75 ± 0.07}  & \underline{49.53 ± 0.02}&	\textbf{51.86 ± 0.01} & 69.98 ± 0.01	&79.75 ± 0.05 & \textbf{47.87 ± 0.14}&	\textbf{50.35 ± 0.14}\\
AlterRec\_{aug} &\textbf{70.82 ± 0.09}	&\textbf{79.84 ± 0.1} & \textbf{49.56 ± 0.06}	& \textbf{51.86 ± 0.07} & \textbf{70.13 ± 0.03}	& \textbf{79.86 ± 0.11} & \textbf{47.87 ± 0.13}	&\underline{50.34 ± 0.15}\\
\bottomrule
\end{tabular}
\end{adjustbox}
\vspace{-0.1in}
\end{table*}

\section{Experiment}


In this section, we conduct comprehensive experiments to validate the effectiveness of AlterRec. In the following, we will  introduce the experimental settings, followed by the results and their analysis.

\subsection{Experimental Settings}
\subsubsection{ Datasets}
\label{sec:datasets}
We adopt two real-world session recommendation datasets including  textual data. \textbf{Homedepot}: It is a private data from the Home Depot that is derived from user purchase logs on its website\footnote{https://www.homedepot.com/}.  
\textbf{Amazon-M2}~\cite{jin2023amazon}: It's a multilingual dataset.
For the purpose of this study, which does not focus on multilingual data, we extracted unilingual sessions to create individual datasets for three languages: Spanish, French, and Italian. They are denoted as \textbf{Amazon-Spanish}, \textbf{Amazon-French}, and \textbf{Amazon-Italian}, respectively. 
More details can be found  in Appendix~\ref{sec:app_dataset}.  

\vspace{-0.05in}
\subsubsection{Baselines}
In our study, we refer to our model without augmentation as \textbf{AlterRec} and to the augmented version as \textbf{AlterRec\_aug}. 
We include a range of baseline methods, encompassing both ID-based approaches and those combining ID and textual data.
 Our experiment specifically includes the following methods: \textbf{CORE}~\cite{hou2022core}, \textbf{SASRec}~\cite{kang2018sasrec}, \textbf{BERT4Rec}~\cite{sun2019bert4rec},  \textbf{SR-GNN}~\cite{wu2019srgnn}, and \textbf{HG-GNN}~\cite{pang2022hggnn} as ID-based methods. Text-integrated methods include, \textbf{LLM2BERT4Rec}~\cite{harte2023llm2bert4rec},\textbf{UniSRec}~\cite{hou2022unisec}, \textbf{FDSA}~\cite{zhang2019fdsa}, and \textbf{S³-Rec}~\cite{zhou2020s3rec}. 
Notably, \textbf{UniSRec (FHCKM)} refers to the UniSRec model pretrained on the FHCKM dataset~\cite{hou2022unisec,ni2019justifying}, as used in the original UniSRec paper, and \textbf{UniSRec} in this work denotes the model pretrained on our datasets, namely Homdepot and three Amazon-M2 datasets. LLM2BERT4Rec incorporates BERT4Rec as one of the backbone models. Additionally, we experiment with another ID-based backbone SASRec as deonted by \textbf{LLM2SASRec}.  To ensure a fair comparison, each baseline method employs the same input features as AlterRec, specifically the sentence embeddings generated by Sentence-BERT. An exception is {UniSRec (FHCKM)}, a pretrained model with fixed dimension sizes.

\vspace{-0.05in}
\subsubsection{Settings}
Empirically, for Homedepot dataset, 
we use the mean function to generate  ID session embedding and Transformer to generate  text session embedding. For Amazon-M2, we use Transformer to generate both ID and text session embeddings.
 To evaluate the model performance, we use two widely adopted metrics Hits@N and NDCG@N, and N is set to be 10 and 20.
Higher scores of these metrics indicate better performance. 
We set the parameters as follows: 
$m_{random} = 2, m_{gap} =2, m_{max} = 30$, $\alpha=0.5$,  $\beta=0.5$, $p=5$. 
Additionally, for the Homedepot dataset, we set  $k_1=6$, $k_2=2000$, 
and for the three Amazon-M2 datasets, $k_1=20$,  $k_2=20000$ are used. More details are given in Appendix~\ref{sec:app_setting}.

\subsection{Performance Comparison}
The  comparison results are presented in Table~\ref{table:mainresult1}.
Note that HG-GNN~\cite{pang2022hggnn}  incorporates both user and item data in its model. However, the Amazon-M2 dataset lacks user information. Consequently, it is not feasible to obtain results for HG-GNN which are denoted as``N/A''. We summarize our observations as follows.

\begin{table}[!ht]
\centering

\caption{Ablation study on key components. Reported results are mean value over three seeds.  }
\label{table:ablation_study}
\begin{adjustbox}{width =0.47 \textwidth}
\begin{tabular}{c|cc|cc}

\toprule
 & \multicolumn{2}{c|} {Homedepot} &  \multicolumn{2}{c} {Amazon-French}   \\

Methods & Hits@10 & Hits@20 & Hits@10 & Hits@20  \\
  \midrule
AlterRec&	\textbf{38.25} &	\textbf{46.31}  & \textbf{70.61} & \textbf{79.75} \\
AlterRec\_random	& 37.41 &	45.41  & 70.46 &	79.64  \\
AlterRec\_w/o\_{text} & 35.64 	&42.95 & 68.26	&77.23\\
AlterRec\_w/o\_{ID}  &	30.05  &	38.73   &66.96	&76.85\\
\bottomrule
\end{tabular}
\end{adjustbox}
\vspace{-0.1in}
\end{table}

\begin{compactenum}[\textbullet]
    \item 
    Alter\_aug consistently outperforms other baseline models across a range of datasets, with AlterRec often achieving the second-best performance, hightlighting the effectiveness of our alternative training strategy. Moreover, it demonstrates that integrating augmentation data can further enhance performance.
    Although UniSRec and FDSA exhibit strong performance in some cases, they do not consistently excel across all metrics. In contrast, AlterRec maintains a balanced and superior performance in both Hits@N and NDCG@N. For instance, AlterRec shows about a 10\% improvement over UniSRec based on NDCG@10 and NDCG@20 on Amazon-M2 datasets. Additionally, it achieves approximately 19\% and 2\% improvements over FDSA based on Hits@10 and Hits@20 on the Homedepot and Amazon-M2 datasets respectively.

    \item 
    When comparing models that incorporate text data with those solely based on IDs, it's observed that models including text data typically demonstrate better performance. For example, AlterRec, along with UniSRec and FDSA, generally outperform ID-based models. This indicates that text information could offer complementary information, thereby enhancing overall performance.
    
\end{compactenum}

\subsection{Ablation Study}
\label{sec:ablation_study}
In this subsection, we evaluate the effectiveness of key components in our model:  the hard negative samples and the ID and text uni-modal networks.
The results of our ablation study are detailed in Table~\ref{table:ablation_study}. Our model variants are denoted as follows: 
"AlterRec\_random" for training with random negative samples, "AlterRec\_w/o\_text" for the model excluding the text uni-modal network, and "AlterRec\_w/o\_ID" for the model without the ID uni-modal network. Notably, AlterRec\_random uses the same number of negative sample with AlterRec. Furthermore, 
AlterRec\_w/o\_text and AlterRec\_w/o\_ID are trained exclusively on a single modality, either ID or text.

The results in the Table~\ref{table:ablation_study} indicate that 
employing random negative samples hurts performance. Notably, using random negatives behaves like the independent training, lacking interaction between the two modalities. This finding highlights the effectiveness of  AlterRec over independent training. Its superior performance is likely due to the use of hard negative samples, which facilitates the learning between the two uni-modal networks. Furthermore, AlterRec significantly outperforms model variants that rely only on ID information, i.e., AlterRec\_w/o\_text.
For instance, AlterRec achieves improvements of 7.82\% and 3.26\% in terms of Hits@20 on the Homedepot and Amazon-French datasets, respectively. These results demonstrate AlterRec's superior ability to integrate text information, highlighting its advantage over naive fusion methods.

\begin{figure}[t]
\begin{center}
 \centerline{
{\subfigure[Homdepot]
{\includegraphics[width=0.5\linewidth]{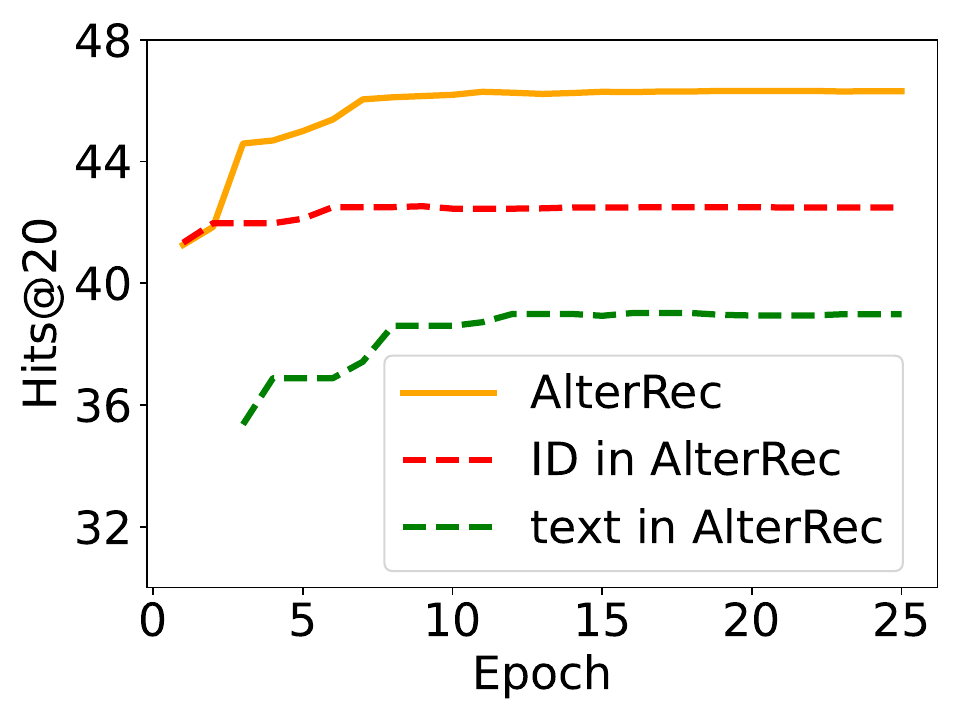} }}

{\subfigure[Amazon-French]
{\includegraphics[width=0.5\linewidth]{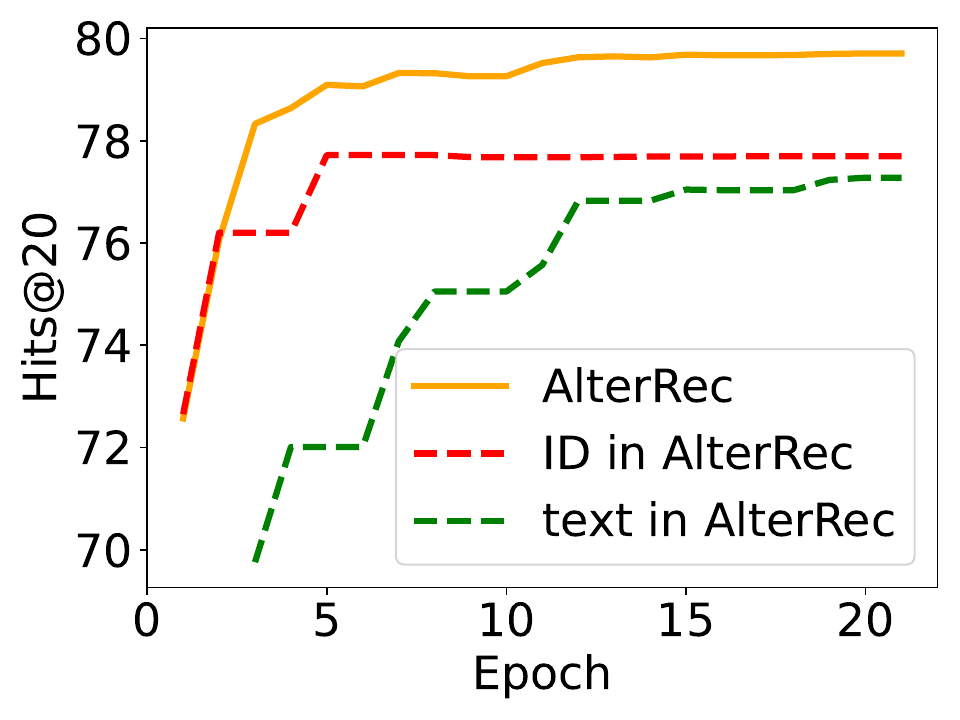} }}
}
\vspace{-0.15in}
\caption{   
Test performance across each epoch during the alternative training.}

\label{fig:ablation}
\end{center}
\vspace{-0.3in}
\end{figure}

Additionally, we present the performance of AlterRec in Figure~\ref{fig:ablation}, including the individual performance of the ID and text components within AlterRec across epochs. These components are denoted as "ID in AlterRec" and "text in AlterRec", respectively. The overall performance of AlterRec is based on the score 
$y_{\mathbf{s},i}$ in Eq.~(\ref{eq:final_score}). The performance of "ID in AlterRec" and "text in AlterRec" are derived from the scores $y^{ID}_{\mathbf{s},i}$ and $y^{text}_{\mathbf{s},i}$ within $y_{\mathbf{s},i}$. 
Figure~\ref{fig:ablation} demonstrates that both ID and text components are effectively trained in our model. It showns that the text modality in AlterRec is well-trained, and crucially,  AlterRec does not exhibit the imbalance issue commonly associated with naive fusion.


\begin{figure}[t]
\begin{center}
 \centerline{
{\subfigure[Homdepot]
{\includegraphics[width=0.5\linewidth]{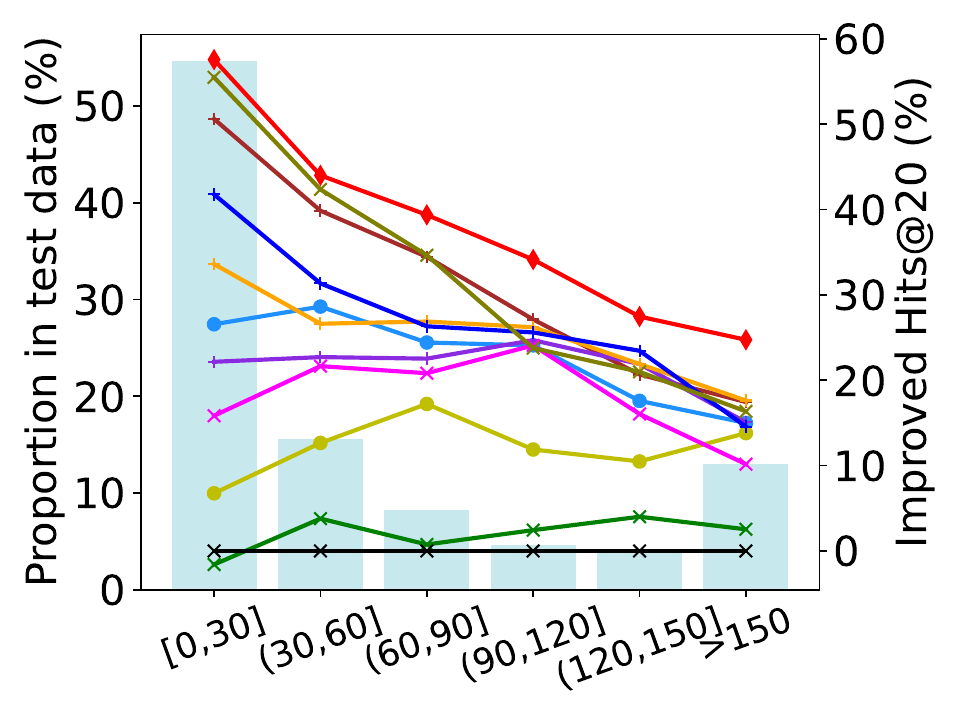} }}

{\subfigure[Amazon-French]
{\includegraphics[width=0.5\linewidth]{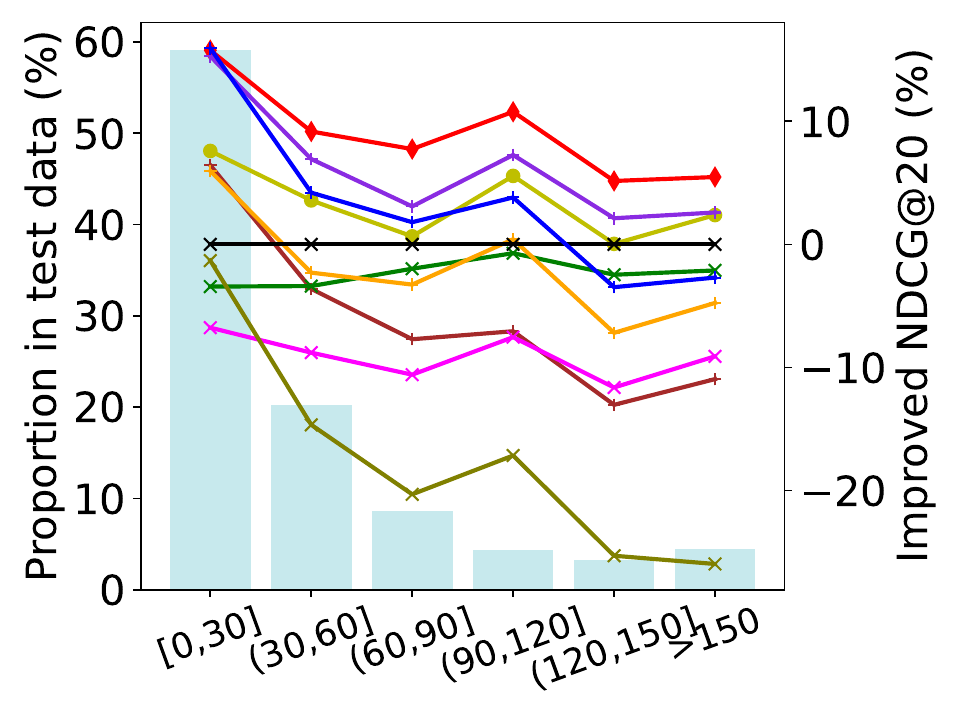} }}
}
\centerline{
\includegraphics[width=0.98\linewidth]{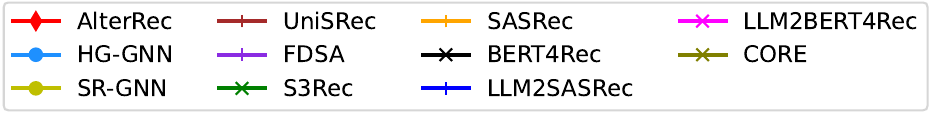}
}
\caption{Performance comparison w.r.t. long-tail items on the Homedepot and Amazon-French datasets.  The bar graph depicts the proportion of interactions in the test data for each group. The line chart illustrates the improvement ratios for Hits@20 and NDCG@20 relative to BERT4Rec.}

\label{fig:long_tail}
\end{center}
\vspace{-0.4in}
\end{figure}
\subsection{  Performance on Long-tail Items }

Textual data offers valuable semantic information that can be  used to enhance long-tail items in session-based recommendation.
To validate this, we divide the test data into groups based on the popularity of the ground-truth item in the training data. We then compare the performance of various methods in each group against the ID-based method BERT4Rec. The comparative result is presented in Figure~\ref{fig:long_tail}, where we also show the proportion of each group.
This figure reveals that a majority of items have sparse interactions (long-tail items). In most cases, AlterRec outperforms other baselines particularly on long-tail items. For instance, AlterRec achieves the best performance in the [0,30] group on the Homedepot and Amazon-French. 
It indicates that AlterRec effectively captures textual information, enhancing its performance on long-tail items. 

\subsection{Parameter Analysis}
\label{sec:parameter_analysis}

\begin{figure}[t]
\begin{center}
 \centerline{
{\subfigure[]
{\includegraphics[width=0.5\linewidth]{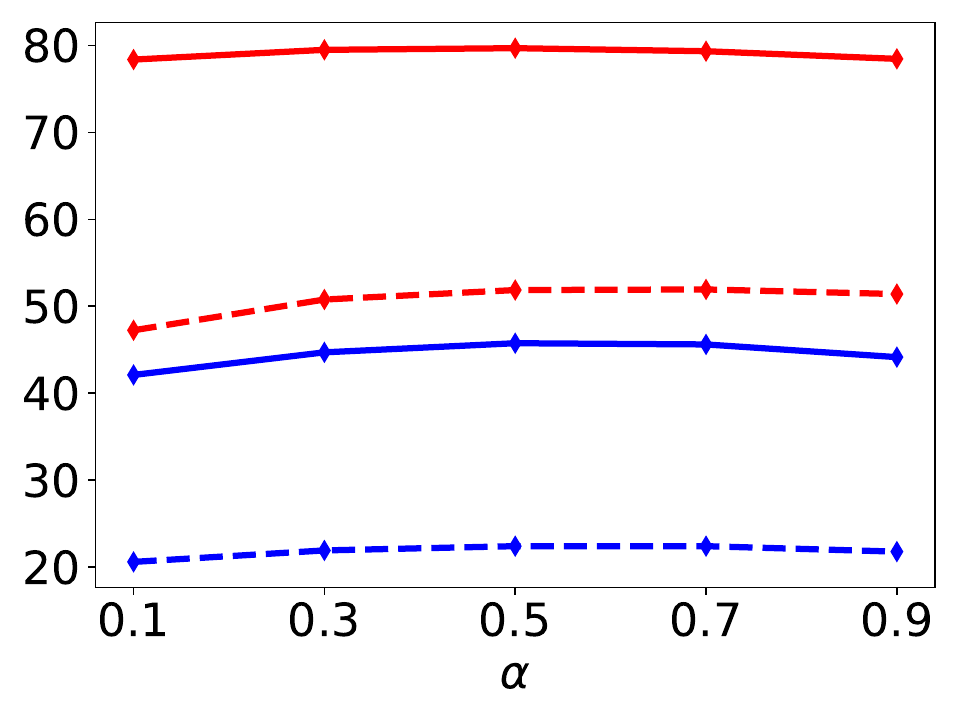} }}

{\subfigure[]
{\includegraphics[width=0.5\linewidth]{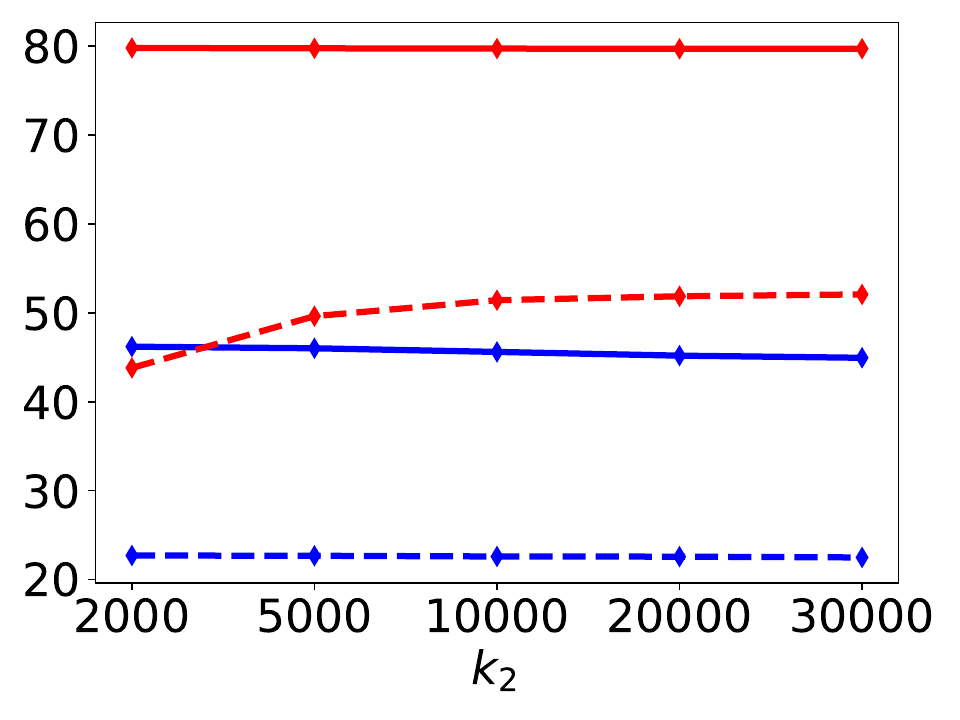} }}
}
\centerline{
\includegraphics[width=0.98\linewidth]{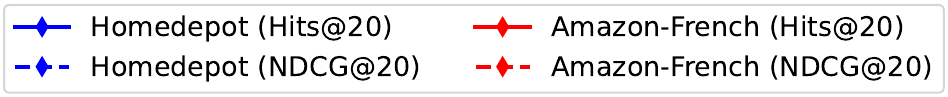}
}
\vspace{-0.1in}
\caption{Performance of AlterRec by varying the hyper-parameter $\alpha$ and $k_2$.}

\label{fig:parameter_analysis}
\end{center}
\vspace{-0.3in}
\end{figure}

In this subsection, we analyze the sensitivity of two key hyper-parameters: the parameter $\alpha$ which adjusts the contribution of ID and text scores in Eq.~(\ref{eq:final_score}), and the end index $k_2$ used for selecting hard negative samples as discussed in Section~\ref{sec:negative_sample}. We explore how these parameters influence the performance  by varying their values across different scales on two datasets, Homedepot and Amazon-French.
The results for Hits@20 and NDCG@20 are presented in Figure~\ref{fig:parameter_analysis}. 
Regarding $\alpha$, we note a similar trend across both datasets with relatively stable performance. An increase in performance is observed as 
$\alpha$ rises from 0.1 to 0.5, followed by a decrease when 
$\alpha$ is increased from 0.5 to 0.9. This pattern suggests that an $\alpha$ value of 0.5 typically yields the best performance, indicating equal contributions from ID and text.
For $k_2$, there is an increasing trend in NDCG@20 on Amazon-French and a decreasing trend in Hits@20 on Homedepot as 
 $k_2$ increases, and the overall performance remains stable. This indicates that the Amazon-French dataset may benefit from relatively more hard negative samples, whereas the Homedepot dataset does not require as many.


\section{Conclusion}

In this work, we explore an effective method for combining ID and text information in session-based recommendation. We have identified an imbalance issue in the widely-used naive fusion framework, which leads to insufficient integration of text information. To address this, we introduce a novel approach AlterRec employing the alternative training strategy that enables implicit interactions between ID and text, thereby facilitating their mutual learning and  enhancing overall performance.
Specifically, we develop separate uni-modal networks for ID and text to capture their respective information. By employing hard negative samples and augmented training samples from one network to train the other, we facilitate the exchange of information between the two, leading to improved overall performance. 
The effectiveness of AlterRec is validated through extensive experiments on various datasets against state-of-the-art baselines, demonstrating its superiority in session recommendation scenarios. 
In the future, we plan to investigate more advanced models, such as LLaMA, as the text encoders in AlterRec.

\bibliographystyle{ACM-Reference-Format}
\bibliography{main}

\newpage
\appendix
\begin{figure*}[t]
\begin{center}
 \centerline{
{\subfigure[NFRec]
{\includegraphics[width=0.33\linewidth]{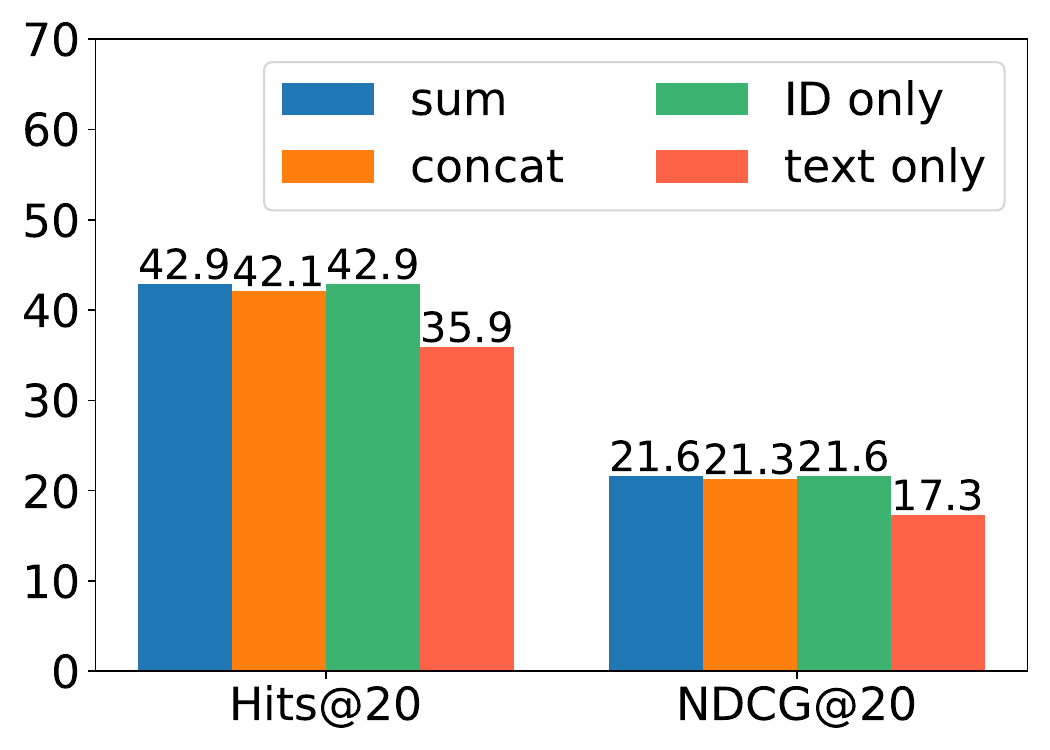} }}

{\subfigure[UnisRec]
{\includegraphics[width=0.33\linewidth]{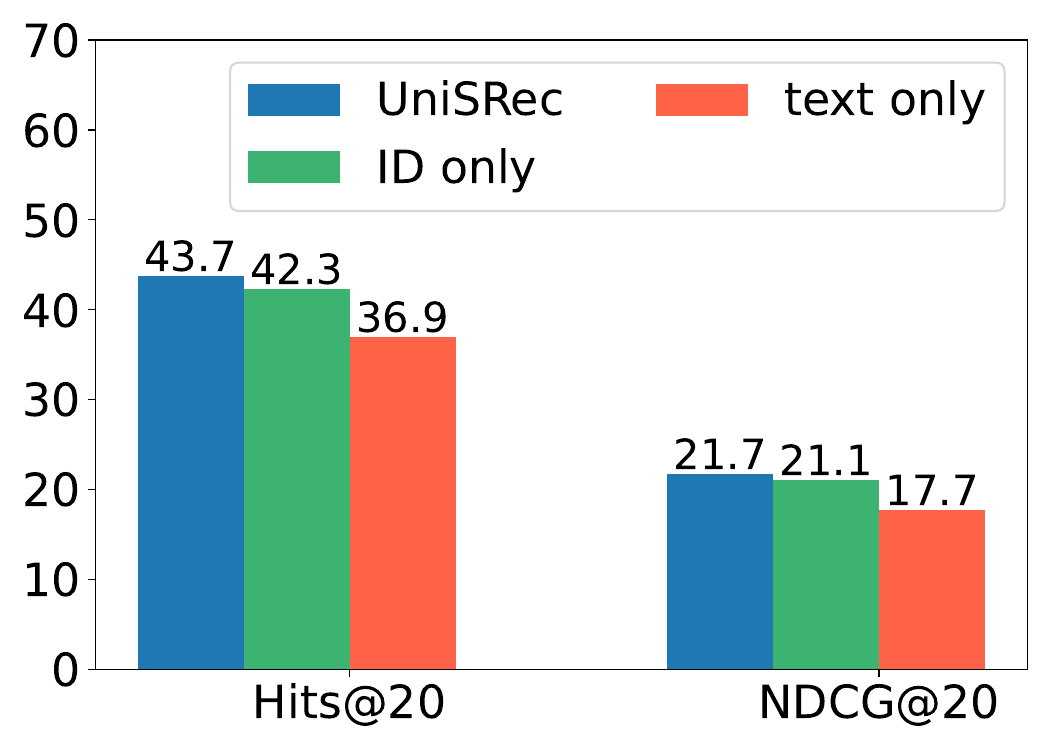} }}

{\subfigure[FDSA]
{\includegraphics[width=0.33\linewidth]{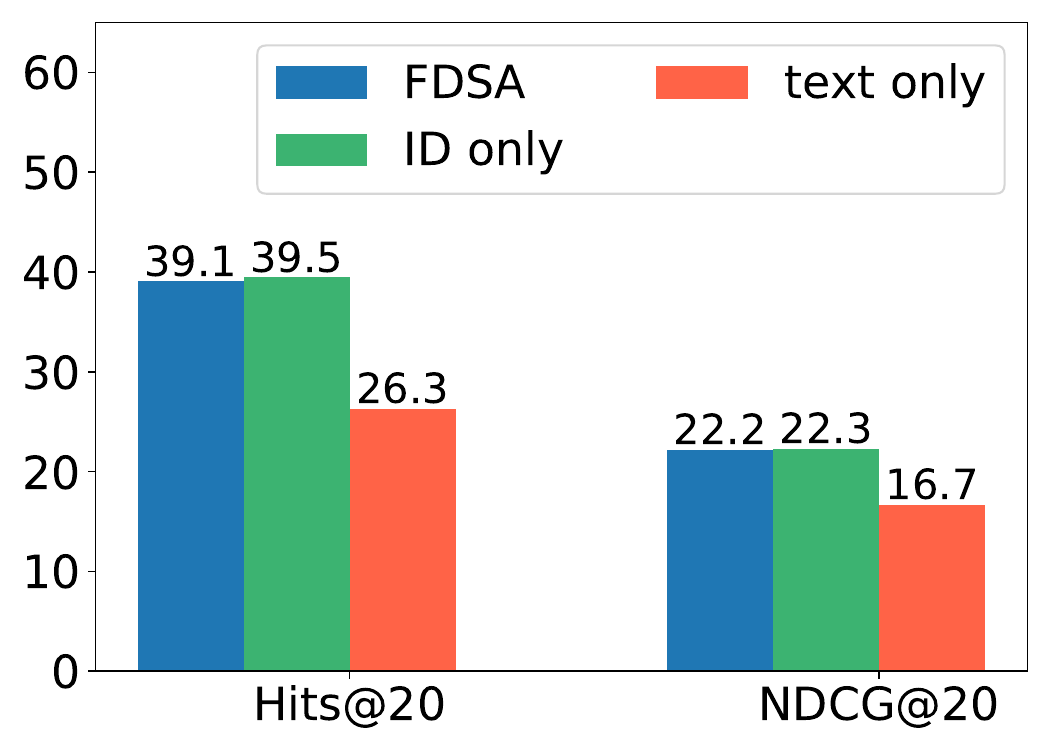} }}

}

\caption{Session recommendation results (\%) on the Homdepot dataset. We compare the models combing ID and text against models trained independently on either ID or text information alone. 
}

\label{fig:app_prili_homedepot}
\end{center}
\end{figure*}

\begin{figure}[t]
\begin{center}
 \centerline{
{\subfigure[Homdepot]
{\includegraphics[width=0.5\linewidth]{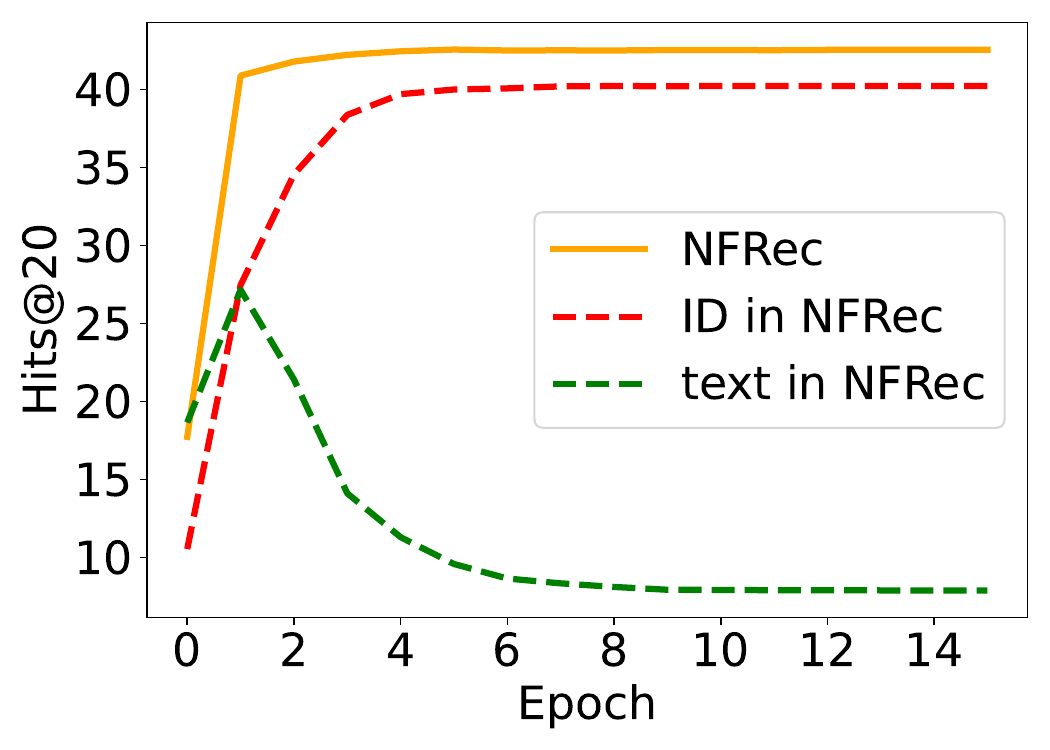} }}

{\subfigure[Homdepot]
{\includegraphics[width=0.5\linewidth]{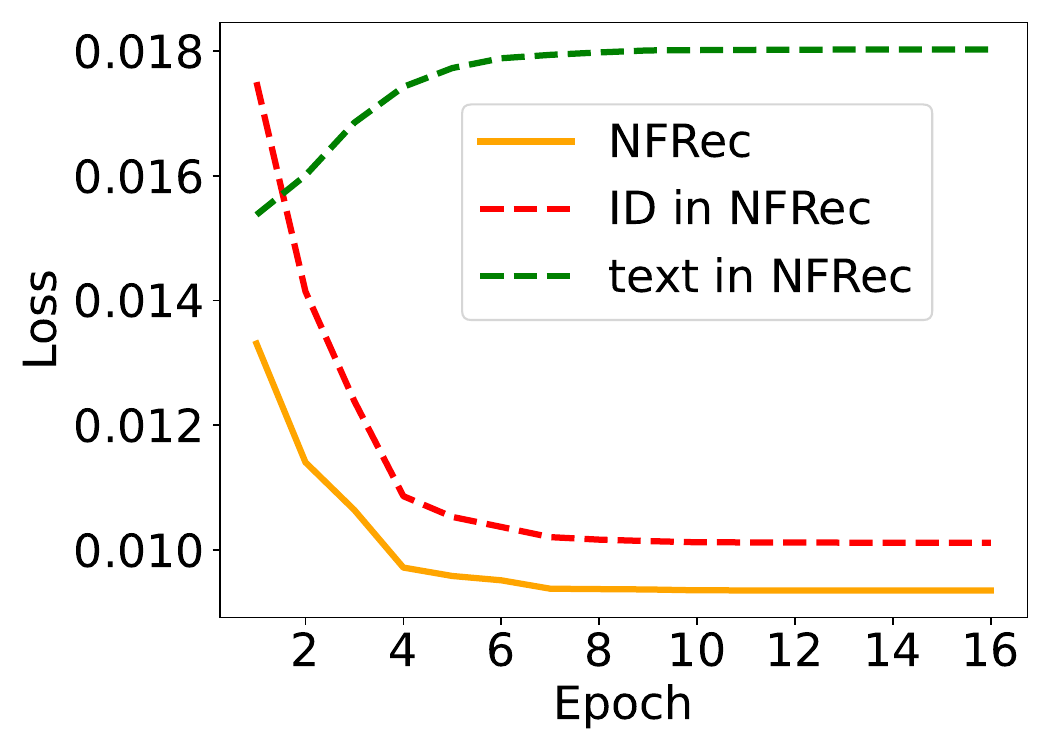} }}
}

\caption{Test performance in terms of Hits@20 (\%) and  training loss comparison on the Homdepot dataset.  }

\label{fig:app_understand}
\end{center}

\end{figure}

\begin{table}[!ht]
\centering
\caption{Data statistic of the session datasets. The Amazon-M2 datasets don't involve  users. \#Train, \#Val, and \#Test denote the number of sessions in the train, validation, and test.  }
\label{table:data}
\begin{tabular}{cccccc}
\toprule
Dataset &\#User& \#Item &\#Train &\#Val &\#Test\\
\midrule
Homedepot& 145,750 & 39,114	& 182,575&	2,947	&5,989\\
Amazon-Spanish& - &38,888&75,098 &	7,900	&6,237 \\
Amazon-French& - & 40,258 & 96,245 &	10,507&	8,981\\
Amazon-Italian &- &45,559 & 102,923	&11,102	&10,158 \\
\bottomrule
\end{tabular}
\end{table}

\section{Implementation of Naive Fusion}
\label{sec:implementation_naive_fusion}
In this section, we give more details of 
the   naive fusion  methods in section~\ref{sec:joint_vs_independent}. We explore three approaches: our own implementation NFRec, UnisRec~\cite{hou2022unisec}, and FDSA~\cite{zhang2019fdsa}, with detains provided in the following.
\begin{compactenum}[\textbullet]
\item \textbf{NFRec}:  It consists of several key components. We give more details of these components. \textbf{ID and text encoder}: We employ the same ID and text encoder as AlterRec which is introduced in section~\ref{sec:id_encoder} and section~\ref{sec:text_encoder}, respectively. Through these two encoders, we obtain the item-level ID embedding $\mathbf{X}$ and text embeddings $\mathbf{H}$. 
\textbf{Fusion operation}: We  fuse the ID and text item embeddings to form a final embedding $\mathbf{Z}$ via summation or concatenation to as mentioned in section~\ref{sec:joint_vs_independent}.
\textbf{Scoring function}: For a given session $\mathbf{s}$, we apply the mean function based on the fused item embedding  to get the session embedding $\mathbf{q}_{\mathbf{s}}=g_{mean}(\mathbf{s}, \mathbf{Z})$, and then
we use the vector multiplication between session embedding and the candidate item's fused embedding to get the score $y_{\mathbf{s}, i} = \mathbf{Z_i}^{T}\mathbf{q}_{\mathbf{s}}$.
\textbf{Loss function}: We use the cross entropy as the loss function, which follows similar form with Eq.~(\ref{eq:id_loss}). 

\item \textbf{UniSRec }~\cite{hou2022unisec}: The model employs the same ID encoder as NFRec, which utilizes a learnable embedding for the ID representation. For text encoder, it leverages a language model enhanced by the proposed adaptor to extract textual information. After pretraining the adaptor using two contrastive loss functions, it merges the ID and text embeddings through summation. UniSRec adopts the same cross-entropy loss function as used in NFRec. We use the official code of UnisRec ~\footnote{https://github.com/RUCAIBox/UniSRec/tree/master} as the implementation.
\item \textbf{FDSA}~\cite{zhang2019fdsa}: The ID encoder generates ID embeddings using learnable embeddings. The text encoder employs a MLP and an attention mechanism to produce text embeddings. The Transformer is applied to items within a session to create ID and text session embeddings, which are then concatenated to form the final session embedding. Similar with NFRec and UniSRec, FDSA utilizes cross-entropy as the loss function. For the implementation of FDSA, we utilize code from the UniSRec's repository, which includes the  implementation details for FDSA.
\end{compactenum}

\section{More Details when exploring NFRec}
\label{sec:app_explore_naive_fusion}
In this section, we give more details for the exploration conducted in section~\ref{sec:explore_naive_fusion}.
We elucidate the process of dividing the NFRec into its ID and text components, and describe how we evaluate the performance and obtain the loss of "ID in NFRec" and "text in NFRec." Details on the implementation of the NFRec are provided in the Appendix~\ref{sec:implementation_naive_fusion}.

For any given item $i$, we derive the ID embedding $\mathbf{X}_i$ and text embedding $\mathbf{H}_i$ from the corresponding ID and text encoders. These two embeddings are then concatenated to form a final embedding $\mathbf{Z}_i = [\mathbf{X}_i, \mathbf{H}_i ]$. For a session $\mathbf{s} = \{ s_1, s_2, ..., s_n\}$, the session embedding  is obtained by applying the mean function to the final embeddings of the items within session $\mathbf{s}$: $\mathbf{q}_{\mathbf{s}}=g_{mean}(\mathbf{s}, \mathbf{Z})$. This session embedding is represented as a concatenation of two parts derived from the ID and text embeddings, respectively:
 \begin{align}
     \nonumber \mathbf{q}_{\mathbf{s}} &= [ \mathbf{q}^{ID}_{\mathbf{s}}, \mathbf{q}^{text}_{\mathbf{s}}] \\
     &=[\frac{1}{|\mathbf{s}|}\sum_{s_i \in \mathbf{s}} \mathbf{X}_{s_i},  \frac{1}{|\mathbf{s}|}\sum_{s_i \in \mathbf{s}} \mathbf{H}_{s_i}]
 \end{align}
The relevance score between a session and an item is then decomposed into two parts:
 \begin{align}
 \label{eq:score_decom}
     \nonumber y_{s,i} &= \mathbf{Z}^T_i \mathbf{q}_s \\
     &= [\mathbf{X}_i, \mathbf{H}_i]^T [ \nonumber \mathbf{q}^{ID}_{\mathbf{s}}, \mathbf{q}^{text}_{\mathbf{s}}] \\
    \nonumber  &=\mathbf{X}^T_i\mathbf{q}^{ID}_{\mathbf{s}} + \mathbf{H}^T_i\mathbf{q}^{text}_{\mathbf{s}} \\
     &= y^{ID}_{s,i} + y^{text}_{s,i}
 \end{align}
Thus, the relevance score in NFRec can be decomposed as  the summation of the ID and text scores. Accordingly, we 
evaluate the performance and obtain the loss of ``NFRec'', ``ID in NFRec'' and ``text in NFRec'' based on $y_{s,i}$,  $y^{ID}_{s,i}$ and $ y^{text}_{s,i}$ in Eq.~(\ref{eq:score_decom}), respectively.  
For the loss function, the cross-entropy is employed. 

\section{Additional Results in Preliminary Study}
\label{sec:app_additional_result}

Additional results on the Homedepot dataset for investigations in sections~\ref{sec:joint_vs_independent} and \ref{sec:explore_naive_fusion} are displayed in \figurename~\ref{fig:app_prili_homedepot} and \figurename~\ref{fig:app_understand}, respectively. These figures indicate a trend similar to that observed with the Amazon-French dataset. Specifically, \figurename~\ref{fig:app_prili_homedepot} reveals that models trained independently on ID data can achieve performance comparable to, or even surpassing, that of naive fusion methods. Furthermore, models relying solely on text information tend to perform the worst. In \figurename~\ref{fig:app_understand}, it is observed that the ID component dominates the performance and loss. These findings are consistent with observations made with the Amazon-French dataset, suggesting that the phenomenon identified in observations 1 and 2 in section~\ref{sec:joint_vs_independent}, as well as the imbalance issue in NFRec, may be prevalent across various datasets.

\section{More Details in Transformer}
\label{sec:app_masking_matrix}

This section provides additional details on the Transformer in section~\ref{sec:id_score}.
We first use positional embeddings $\mathbf{P} \in \mathbb{R}^{n \times d}$ to indicate the position of each item in the sequence:
\begin{equation}
\mathbf{E} = 
\begin{bmatrix}
\mathbf{X}_{s_1} + \mathbf{P}_1 \\
\mathbf{X}_{s_2} + \mathbf{P}_2 \\
\vdots \\
\mathbf{X}_{s_n} + \mathbf{P}_n \\
\end{bmatrix}
\end{equation}
Next, we apply a self-attention layer~\cite{vaswani2017attention}  to aggregate information from other items. The process is formally defined as:
\begin{equation}
\hat{\mathbf{E}}= \text{softmax} \Bigg(  \frac{\mathbf{Q}\mathbf{K}^T}{\sqrt{d}} \odot \mathbf{M}\Bigg)\mathbf{V}
\end{equation}
Here, $\mathbf{Q} = \mathbf{E}\mathbf{W}^Q, \mathbf{K} = \mathbf{E}\mathbf{W}^K, \mathbf{V}= \mathbf{E}\mathbf{W}^V$ are the queries, keys, and values respectively, with $\mathbf{W}^Q, \mathbf{W}^K, \mathbf{W}^V \in \mathbb{R}^{d\times d}$ being three learnable parameters. The factor $\sqrt{d}$ is used to normalize the values, particularly beneficial when the dimension size is large. Originally, self-attention computes a weighted sum of values. However, considering the sequential nature the interaction data, we employ a masking matrix $\mathbf{M} \in \mathbb{R}^{n \times n}$ to prevent the model from accessing future information. In this matrix, $\mathbf{M}_{ij} = 1$ if $i < j$, and  $\mathbf{M}_{ij} = 0$, otherwise, with $\odot$ representing element-wise multiplication.


\section{Alternative training algorithm}
\label{sec:app_algorithm}
We present the pseudo code of the alternative training algorithm in Algorithm~\ref{alg:alter_train}.  The parameters within the ID and text uni-modal networks are denoted as $\theta^{ID}$ and $\theta^{text}$, respectively.
Initially, as indicated in line 1, 
both networks are randomly initialized.
In the early stages of training, both networks are trained with random negative samples, as indicated in line 2-7. It's because the embedding learned in the early stage are of lower quality and might not be able to provide useful information. As training progresses, we shift towards employing hard negative samples. At first, 
the ID unimodal network is trained using predictions from the text unimodal network, as described in lines 9 to 11. After $m_{gap}$ 
  epochs, training shifts to the text unimodal network, utilizing predictions from the ID unimodal network, as indicated in lines 12 to 15. Subsequently, training alternates back to the ID network. This cycle continues until convergence is achieved for both networks.  Notably, for AlterRec\_{aug}, we replace the loss function in line 10 and 13 as Eq.~(\ref{eq:id_loss_aug}) and  Eq.~(\ref{eq:text_loss_aug}) respectively.

\begin{algorithm}

\caption{Alternative Training }
\label{alg:alter_train}
\begin{algorithmic}[1] %
\REQUIRE User-item interaction set $\mathcal{S}$, epoch number using random negatives $m_{random}$, maximum epoch number $m_{max}$, gap epoch number $m_{gap}$
\ENSURE Converged models  $\theta^{ID}$, $\theta^{text}$
\STATE Random initialize two uni-modal networks $\theta^{ID}$, $\theta^{text}$
\FOR{i = 1, 2, \dots, $m_{random}$ } 
\STATE Train  $\theta^{ID}$ using random negatives
\ENDFOR
\FOR{i = 1, 2, \dots, $m_{random}$ } 
\STATE  Train $\theta^{text}$ using random negatives
\ENDFOR
\FOR{i = 0, 1,  \dots, $m_{max} - 2*m_{random}$ } 
\IF{$i \mod (2*m_{gap} )  <  m_{gap}$ }
\STATE Compute loss in Eq.~(\ref{eq:id_loss})
\STATE Update $\theta^{ID}: \theta^{ID} \leftarrow \theta^{ID} - \alpha {\nabla} L^{ID} $
\ELSE
\STATE Compute loss in Eq.~(\ref{eq:text_loss}) 
\STATE Update $\theta^{text}: \theta^{text} \leftarrow \theta^{text} - \alpha {\nabla} L^{text} $
\ENDIF
\ENDFOR
\end{algorithmic}
\end{algorithm}

\section{Datasets}
\label{sec:app_dataset}

We provide the data statistics in Table~\ref{table:data}. The \textbf{Homedepot} dataset is a sampled dataset of purchase logs from the Homedepot's website. We include sessions where all items have textual data, i.e., titles, descriptions, and taxonomy. 
Items in each session is interacted by the same user, who may have engaged in several sessions at distinct timestamps. For the purposes of validation and testing, we select the most recent sessions from different users. Specifically, 10\% of these sessions are designated for validation and 20\% for testing, with the remainder allocated to training sessions.
Typically, the sessions in validation appear after those in the training set, and the sessions for testing appear after those in the validation set.
For the three \textbf{Amazon-M2} datasets, since there is no original validation set, we use about 10\% of the training set to create a validation set.


 \section{Experimental Settings}
 \label{sec:app_setting}
 In our experimental setup, we search the learning rate in $\{0.01, 0.001\}$ and dropout in
$\{0.1, 0.3, 0.5\}$, and we set hidden dimension as 300, and number of Transformer layer to be 2, for all  models. The test results we report are based on the model that achieves the best performance during the validation phase.
For text feature extraction in the Homedepot dataset, we utilize Sentence-BERT with the all-MiniLM-L6-v2 model\footnote{https://huggingface.co/sentence-transformers/all-MiniLM-L6-v2}. In contrast, for the three Amazon-M2 datasets, we employ Sentence-BERT with the distiluse-base-multilingual-cased-v1 model\footnote{https://huggingface.co/sentence-transformers/distiluse-base-multilingual-cased-v1}, due to its proficiency in handling multiple languages including Spanish, French, and Italian.
For each item in the Homedepot dataset, we use title, description, and taxonomy as the textual data. For the Amazon-M2 datasets,  we use the title and description as textual data. All baseline methods employ the cross entropy  as loss function and are implemented based on the RecBole~\footnote{https://recbole.io/index.html}. 


\end{document}